\pdfoutput=1

\documentclass[11pt,twoside,a4paper,cmspaper,final,collab]{cms-tdr}
\def\svnVersion{215789}\def\cmsCernNo{2013-152}\def\cmsCernDate{\today}\def\cmsMessage{Published in Physics Letters B as \href{http://dx.doi.org/10.1016/j.physletb.2013.10.017}{\doi{10.1016/j.physletb.2013.10.017.}}}

\begin{document}\cmsNoteHeader{BPH-11-009}

\hyphenation{had-ron-i-za-tion}
\hyphenation{cal-or-i-me-ter}
\hyphenation{de-vices}
\RCS$HeadURL: svn+ssh://svn.cern.ch/reps/tdr2/papers/BPH-11-009/trunk/BPH-11-009.tex $
\RCS$Id: BPH-11-009.tex 209480 2013-10-01 19:04:40Z alverson $
\newlength\cmsFigWidth
\ifthenelse{\boolean{cms@external}}{\setlength\cmsFigWidth{0.85\columnwidth}}{\setlength\cmsFigWidth{0.4\textwidth}}
\ifthenelse{\boolean{cms@external}}{\providecommand{\cmsTop}{top}}{\providecommand{\cmsTop}{top left}}
\ifthenelse{\boolean{cms@external}}{\providecommand{\cmsMiddle}{middle}}{\providecommand{\cmsMiddle}{top right}}
\ifthenelse{\boolean{cms@external}}{\providecommand{\cmsBottom}{bottom}}{\providecommand{\cmsBottom}{bottom}}
\ifthenelse{\boolean{cms@external}}{\providecommand{\cmsLeft}{top}}{\providecommand{\cmsLeft}{left}}
\ifthenelse{\boolean{cms@external}}{\providecommand{\cmsRight}{bottom}}{\providecommand{\cmsRight}{right}}
\providecommand{\cPKst}{\ensuremath{\cmsSymbolFace{K}^\ast}\xspace}
\providecommand{\cPKstz}{\ensuremath{\cmsSymbolFace{K}^{\ast0}}\xspace}
\providecommand{\cPAKstz}{\ensuremath{\overline{\cmsSymbolFace{K}}{}^{\ast0}}\xspace}
\newcommand{\BtoKstmumu}{\ensuremath{\PBz\to\cPKstz \Pgmp \Pgmm}\xspace}
\newcommand{\BtoKstJpsi}{\ensuremath{\PBz\to\cPKstz \cPJgy}\xspace}
\newcommand{\BtoKstpsip}{\ensuremath{\PBz\to\cPKstz \psi'}\xspace}
\newcommand{\BtoKstJpsimumu}{\ensuremath{\PBz\to\cPKstz \cPJgy(\Pgmp \Pgmm})\xspace}
\newcommand{\BtoKstpsipmumu}{\ensuremath{\PBz\to\cPKstz \psi'(\Pgmp \Pgmm)}\xspace}
\newcommand{\BtoKstmumudecay}{\ensuremath{\PBz\to\cPKstz(\PKp \Pgpm) \Pgmp \Pgmm}\xspace}
\newcommand{\BtoKstJpsidecay}{\ensuremath{\PBz\to\cPKstz(\PKp \Pgpm) \cPJgy(\Pgmp \Pgmm)}\xspace}
\newcommand{\BtoKstpsipdecay}{\ensuremath{\PBz\to\cPKstz(\PKp \Pgpm) \psi'(\Pgmp \Pgmm)}\xspace}
\renewcommand{\PaBz}{\ensuremath{\overline{\cmsSymbolFace{B}}{}^{0}}\xspace}

\cmsNoteHeader{BPH-11-009}

\title{Angular analysis and branching fraction measurement of the decay \texorpdfstring{$\PBz\to\cPKstz\Pgmp\Pgmm$}{B0 to K*0 mu+ mu-}}

\date{\today}

\abstract{The angular distributions and the differential branching fraction of the decay
$\PBz \to \PKst{}^0 \Pgmp \Pgmm$ are studied using a data sample
corresponding to an integrated luminosity of 5.2\fbinv collected with the CMS detector at
the LHC in pp collisions at $\sqrt{s} = 7$\TeV.
From more than 400 signal decays, the forward-backward asymmetry of the muons, the $\PKst{}^0$
longitudinal polarization fraction, and the differential branching fraction are determined as a
function of the square of the dimuon invariant mass. The measurements are in good agreement with
standard model predictions.}

\hypersetup{%
pdfauthor={CMS Collaboration},%
pdftitle={Angular analysis and branching fraction measurement of the decay B0 to K*0 mu+ mu-},%
pdfsubject={CMS},%
pdfkeywords={CMS, physics}}

\maketitle

\section{Introduction}
\label{sec:Intro}

It is possible for new phenomena (NP) beyond the standard model (SM) of particle physics to be
observed either directly or indirectly, \ie, through their influence on other physics processes.
Indirect searches for NP generally proceed by comparing experimental results with theoretical
predictions in the production or decay of known particles.
The study of flavor-changing neutral-current decays of b hadrons such as \BtoKstmumu
($\cPKstz$ indicates the $\PKst{}^0$ and charge
conjugate states are implied in what follows, unless explicitly stated otherwise) is particularly
fertile for new phenomena searches, given the modest theoretical uncertainties in the predictions and
the low rate as the decay is forbidden at tree level in the SM\@. On the theoretical side, great
progress has been made since the first calculations of the branching
fraction~\cite{Deshpande:1988mg,Deshpande:1988bd,Lim:1988yu,Grinstein:1988me},
the forward-backward asymmetry of the muons, $A_\mathrm{FB}$~\cite{Ali:1991is},
and the longitudinal polarization fraction of the $\cPKstz$,
$F_L$~\cite{Kruger:1999xa,Kim:2000dq,Yan:2000dc,Aliev:2001fc,Beneke:2001at,Chen:2002zk}.
Robust calculations of these variables~\cite{Bobeth:2010wg, Bobeth:2011nj, Bobeth:2012vn,
Ali:2006ew, Altmannshofer:2008dz, Altmannshofer:2011gn, Jager:2012uw, Descotes-Genon:2013vna} are now available for much
of the phase space of this decay, and it is clear that new physics could give rise to readily observable
effects~\cite{Altmannshofer:2008dz, Melikhov:1998cd, Ali:1999mm, Yan:2000dc, Buchalla:2000sk, Feldmann:2002iw, Hiller:2003js,
Kruger:2005ep, Hovhannisyan:2007pb, Egede:2008uy, Hurth:2008jc, Alok:2009tz, Alok:2010zd, Chang:2010zy, DescotesGenon:2011yn,
Matias:2012xw,DescotesGenon:2012zf}.
Finally, this decay mode is relatively easy to select and reconstruct at hadron colliders.

The quantities $A_\mathrm{FB}$ and $F_L$ can be measured as a function of the dimuon invariant
mass squared $(q^2)$ and compared to SM predictions~\cite{Bobeth:2012vn}.  Deviations from the SM
predictions can indicate new physics.  For example, in the minimal supersymmetric standard model
(MSSM) modified with minimal flavor violation, called flavor blind MSSM (FBMSSM), effects can arise
through NP contributions to the Wilson coefficient $C_7$~\cite{Altmannshofer:2008dz}.  Another NP
example is the MSSM with generic flavor-violating and CP-violating soft SUSY-breaking terms (GMSSM),
in which the Wilson coefficients $C_7$, $C^\prime_7$, and $C_{10}$ can receive
contributions~\cite{Altmannshofer:2008dz}.  As shown there, these NP contributions can dramatically
affect the $A_\mathrm{FB}$ distribution (note that the variable $S^s_6$ defined in
Ref.~\cite{Altmannshofer:2008dz} is related to $A_\mathrm{FB}$ measured in this paper by $S^s_6 =
-{\frac{4}{3}}A_\mathrm{FB}$), indicating that precision measurements of $A_\mathrm{FB}$ can be used
to identify or constrain new physics.

While previous measurements by BaBar, Belle, CDF, and LHCb are consistent with the SM~\cite{BaBar,
Belle, CDF, LHCb}, these measurements are still statistically limited, and more precise
measurements offer the possibility to uncover physics beyond the SM\@.

In this Letter, we present measurements of $A_\mathrm{FB}$, $F_L$, and the differential branching fraction
$\rd{}\mathcal{B}/\rd{}q^2$ from \BtoKstmumu decays, using data collected from pp collisions
at the Large Hadron Collider (LHC) with the Compact Muon Solenoid (CMS) experiment in 2011 at a
center-of-mass energy of 7\TeV.  The analyzed data correspond to an integrated luminosity of
$5.2\pm0.1\fbinv$~\cite{LUMI}.  The $\cPKstz$ is reconstructed through its decay to
$\PKp\Pgpm$ and the $\PBz$ is reconstructed by fitting the two identified muon tracks and the
two hadron tracks to a common vertex. The values of $A_\mathrm{FB}$ and $F_L$ are measured by fitting the
distribution of events as a function of two angular variables: the angle between the positively
charged muon and the $\PBz$ in the dimuon rest frame, and the angle between the kaon and the $\PBz$
in the $\cPKstz$ rest frame.  All measurements are performed in $q^2$ bins from 1 to
$19\GeV^2$.  The $q^2$ bins $8.68<q^2<10.09\GeV^2$ and
$12.90<q^2<14.18\GeV^2$, corresponding to the \BtoKstJpsi and \BtoKstpsip decays
($\psi'$ indicates the \Pgy\ in what follows),
respectively, are both used to validate the analysis, and the former is used to normalize the
branching fraction measurement.

\section{CMS detector}
\label{sec:Detector}

A detailed description of the CMS detector can be found elsewhere~\cite{CMS}. The main detector
components used in this analysis are the silicon tracker and the muon detection systems. The silicon
tracker measures charged particles within the pseudorapidity range $\abs{\eta}<2.4$, where $\eta =
-\ln[\tan(\theta/2)]$ and $\theta$ is the polar angle of the track relative to the
beam direction. It consists of 1440 silicon pixel and 15\,148 silicon strip
detector modules and is located in the 3.8\unit{T} field of the superconducting solenoid.  The
reconstructed tracks have a transverse impact parameter resolution ranging from ${\approx} 100\micron$
to ${\approx} 20\micron$ as the transverse momentum of the track (\pt) increases from 1\GeV to 10\GeV.
In the same \pt regime, the momentum resolution is better than 1\% in the central region, increasing to 2\%
at $\eta \approx 2$, while the track reconstruction efficiency is nearly 100\% for muons with $\abs{\eta}<2.4$
and varies from ${\approx} 95\%$ at $\eta=0$ to ${\approx} 85\%$ at $\abs{\eta}=2.4$ for hadrons.
Muons are measured in the pseudorapidity range $\abs{\eta}<2.4$, with detection planes made using three
technologies: drift tubes, cathode strip chambers, and resistive-plate chambers, all of which are
sandwiched between the solenoid flux return steel plates.  Events are selected
with a two-level trigger system. The first level is composed of custom hardware processors and uses
information from the calorimeters and muon systems to select the most interesting events. The
high-level trigger processor farm further decreases the event rate from nearly 100\unit{kHz} to around
350\unit{Hz} before data storage.

\section{Reconstruction, event selection, and efficiency}
\label{sec:Selection}

The signal (\BtoKstmumu) and normalization/control samples (\BtoKstJpsi and \BtoKstpsip) were
recorded with the same trigger, requiring two identified muons of opposite charge to form a vertex
that is displaced from the pp collision region (beamspot).  The beamspot position and size were
continuously measured from Gaussian fits to reconstructed vertices as part of the online data
quality monitoring.  Five dimuon trigger configurations were used during 2011 data taking with
increasingly stringent requirements to maintain an acceptable trigger rate as the instantaneous
luminosity increased.  For all triggers, the separation between the beamspot and the dimuon vertex
in the transverse plane was required to be larger than three times the sum in quadrature of the
distance uncertainty and the beamspot size.  In addition, the cosine of the angle between the dimuon
momentum vector and the vector from the beamspot to the dimuon vertex in the transverse plane
was required to be greater than 0.9.  More than 95\% of the data were collected with triggers that
required single-muon pseudorapidity of $\abs{\eta(\mu)}<2.2$ for both muons, dimuon transverse momentum of
$\pt(\mu\mu)>6.9\GeV$, single-muon transverse momentum for both muons of
$\pt(\mu)>3.0,$ 4.0, 4.5, 5.0\GeV (depending on the trigger), and the corresponding vertex fit
probability of $\chi^2_\text{prob} > 5\%$, 15\%, 15\%, 15\%.  The remaining data were obtained
from a trigger with requirements of $|\eta(\mu)|<2.5$, $\chi^2_\text{prob}>0.16\%$, and
$\pt(\mu\mu)>6.5\GeV$.
The events used in this analysis passed at least one of the five triggers.

The decay modes used in this analysis require two reconstructed muons and two charged hadrons,
obtained from offline reconstruction.  The reconstructed muons are required to match the muons that
triggered the event readout and to pass several muon identification requirements, namely a track
matched with at least one muon segment, a track fit $\chi^2$ per degree of freedom less than 1.8, at
least 11 hits in the tracker with at least 2 from the pixel detector, and a transverse
(longitudinal) impact parameter less than 3\cm (30\cm).  The reconstructed dimuon system is further
required to satisfy the same requirements as were used in the trigger.  In events where multiple
trigger configurations are satisfied, the requirements associated with the loosest trigger are used.

While the muon requirements are based on the trigger and a CMS standard selection, most of the
remaining selection criteria are optimized by maximizing $S/\sqrt{S+B}$, where $S$ is the expected
signal yield from Monte Carlo (MC) simulations and $B$ is the background estimated from invariant-mass
sidebands in data, defined as ${>}3\sigma_{m(\PBz)}$ and ${<}5.5\sigma_{m(\PBz)}$ from
the $\PBz$ mass~\cite{PDG}, where $\sigma_{m(\PBz)}$ is the average $\PBz$ mass
resolution of $44\MeV$.  The optimization is performed on one trigger sample, corresponding to an
integrated luminosity of 2.7\fbinv, requiring $1.0<q^2<7.3\GeV^2$ or
$16<q^2<19\GeV^2$ to avoid $\cPJgy$ and $\psi'$ contributions.  The hadron tracks are
required to fail the muon identification criteria, and have $\pt(\text{h})>0.75\GeV$ and an extrapolated
distance of closest approach to the beamspot in the transverse plane greater than 1.3 times the sum
in quadrature of the distance uncertainty and the beamspot transverse size.  The two hadrons must
have an invariant mass within 80\MeV of the nominal $\cPKstz$ mass for either the
$\PKp\Pgpm$ or $\text{K}^-\pi^+$ hypothesis.  To remove contamination from $\phi$ decays, the
hadron-pair invariant mass must be greater than 1.035\GeV when the charged \PK\ mass is
assigned to both hadron tracks.  The $\PBz$ candidates are obtained by fitting the four
charged tracks to a common vertex and applying a vertex constraint to improve the resolution of the
track parameters.  The $\PBz$ candidates must have $\pt(\PBz)>8\GeV$,
$|\eta(\PBz)|<2.2$, vertex fit probability $\chi^2_\text{prob} > 9\%$, vertex transverse
separation from the beamspot greater than 12 times the sum in quadrature of the separation
uncertainty and the beamspot transverse size, and $\cos{\alpha_{xy}}>0.9994$, where $\alpha_{xy}$ is
the angle, in the transverse plane, between the $\PBz$ momentum vector and the line-of-flight
between the beamspot and the $\PBz$ vertex.  The invariant mass of the four-track vertex must
also be within 280\MeV of the world-average $\PBz$ mass for either the
$\PKm\Pgpp\Pgmp\Pgmm$ or $\PKp\Pgpm\Pgmp\Pgmm$ hypothesis.  This selection results in an
average of 1.06 candidates per event in which at least one candidate is found.  A single candidate
is chosen from each event based on the best $\PBz$ vertex fit $\chi^2$.

The four-track vertex candidate is identified as a $\PBz \big(\PaBz\big)$
if the $\PKp\Pgpm \big(\PKm\Pgpp\big)$ invariant mass is closest to the nominal
$\cPKstz$ mass.  In cases where both $\PK\Pgp$ combinations are within 50\MeV of the nominal
$\cPKstz$ mass, the event is rejected since no clear identification is possible owing to the
50\MeV natural width of the $\cPKstz$. The fraction of candidates assigned the incorrect
state is estimated from simulations to be 8\%.

From the retained events, the dimuon invariant mass $q$ and its corresponding calculated uncertainty
$\sigma_{q}$ are used to distinguish between the signal and normalization/control samples.  The
\BtoKstJpsi and \BtoKstpsip samples are defined as $m_{\cPJgy}-5\sigma_{q} < q <
m_{\cPJgy}+3\sigma_{q}$ and $\abs{q - m_{\psi'}} < 3\sigma_{q}$, respectively, where
$m_{\cPJgy}$ and $m_{\psi'}$ are the world-average mass values.  The asymmetric selection of the
$\cPJgy$ sample is due to the radiative tail in the dimuon spectrum, while the smaller signal in the
$\psi'$ mode made an asymmetric selection unnecessary.  The signal sample is the complement of the
$\cPJgy$ and $\psi'$ samples.

The global efficiency, $\epsilon$, is the product of the acceptance and the trigger, reconstruction,
and selection efficiencies, all of which are obtained from MC simulations.  The pp collisions are
simulated using \PYTHIA~\cite{Pythia} version 6.424, the unstable particles are decayed by
\EVTGEN~\cite{EvtGen} version 9.1 (using the default matrix element for the signal),
and the particles are traced through a detailed model of the detector with \GEANTfour~\cite{Geant4}.
The reconstruction and event selection for the
generated samples proceed as for the data events.  Three simulation samples were created in which
the $\PBz$ was forced to decay to \BtoKstmumudecay, \BtoKstJpsidecay, or \BtoKstpsipdecay.
The acceptance is calculated as the fraction of events passing the single-muon cuts of
$\pt(\mu)>2.8\GeV$ and $\abs{\eta(\mu)}<2.3$ relative to all events with a $\PBz$ in the event
with $\pt(\PBz)>8\GeV$ and $\abs{\eta(\PBz)}<2.2$.  The acceptance is obtained from the
generated events before the particle tracing with \GEANTfour.  To obtain the reconstruction and
selection efficiency, the MC simulation events are divided into five samples, appropriately sized to
match the amount of data taken with each of the five triggers.  In each of the five samples,
the appropriate trigger and matching offline event selection is applied.  Furthermore, each of the
five samples is reweighted to obtain the correct distribution of pileup events (additional pp
collisions in the same bunch crossing as the collision that produced the $\PBz$ candidate),
corresponding to the data period during which the trigger was active.
The reconstruction and selection efficiency is the ratio of the number events that pass all the
selections and have a reconstructed $\PBz$ compatible with the generated $\PBz$ in the
event relative to the number of events that pass the acceptance criteria.  The compatibility of
generated and reconstructed particles is enforced by requiring the reconstructed $\PKp$,
$\Pgpm$, $\Pgmp$, and $\Pgmm$ to have $\sqrt{(\Delta \eta)^2 + (\Delta \varphi)^2} < 0.3$ for
hadrons and 0.004 for muons, where $\Delta \eta$ and $\Delta \varphi$ are the differences in $\eta$
and $\varphi$ between the reconstructed and generated particles, and $\varphi$ is the azimuthal
angle in the plane perpendicular to the beam direction.  The efficiency and purity of this
compatibility requirement are greater than 99\%.

\section{Analysis method}
\label{sec:Analysis}

\begin{figure}[bht]
  \begin{center}
    \includegraphics[width=\cmsFigWidth]{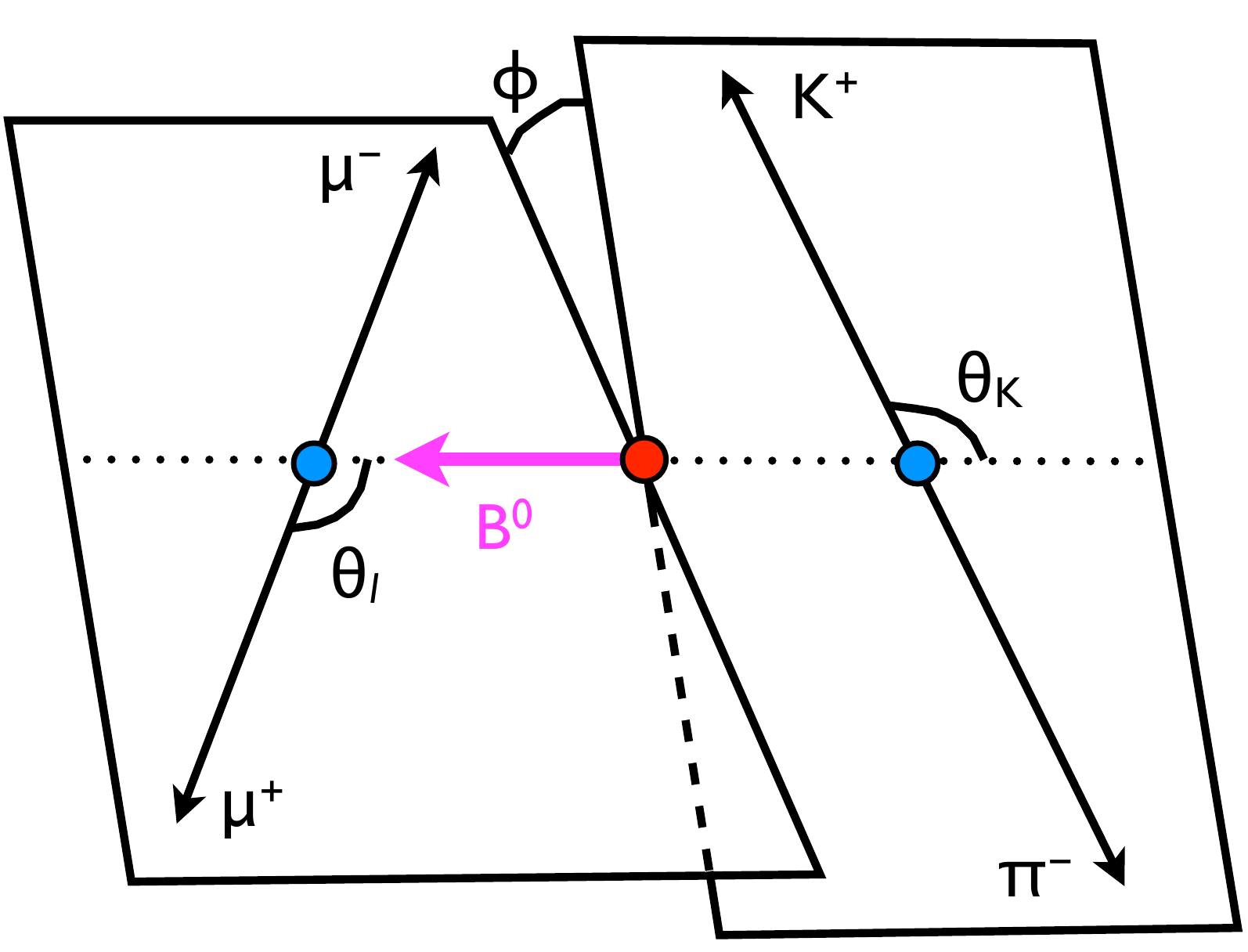}
    \caption{Sketch showing the definition of the angular observables for the decay \BtoKstmumudecay.}
    \label{fig:ske}
  \end{center}
\end{figure}

The analysis measures $A_\mathrm{FB}$, $F_L$, and $\rd{}\mathcal{B}/\rd{}q^2$ of the decay
\BtoKstmumu as a function of $q^2$.  Figure~\ref{fig:ske} shows the relevant angular observables
needed to define the decay: $\theta_\PK$ is the angle between the kaon momentum and the direction
opposite to the $\PBz$ $\big(\PaBz\big)$ in the $\cPKstz$
$\big(\cPAKstz\big)$ rest frame, $\theta_l$ is the angle between the
positive (negative) muon momentum and the direction opposite to the $\PBz$
$\big(\PaBz\big)$ in the dimuon rest frame, and $\phi$ is the angle between the
plane containing the two muons and the plane containing the kaon and pion.
Since the extracted angular parameters $A_\mathrm{FB}$ and $F_L$ and the acceptance times efficiency do not
depend on $\phi$, $\phi$ is integrated out.  Although the $\PKp\Pgpm$ invariant mass must be
consistent with a $\cPKstz$, there can be contributions from a spinless (S-wave)
$\PKp\Pgpm$ combination~\cite{Becirevic:2012dp,Matias:2012qz,Blake:Swave}.
This is parametrized with two terms related to the S-wave fraction,
$F_S$, and the interference amplitude between the S-wave and P-wave decays, $A_S$.  Including this
component, the angular distribution of \BtoKstmumu can be written as~\cite{Blake:Swave}:
\ifthenelse{\boolean{cms@external}}{
\begin{multline}\label{eq:angALL}
  \frac{1}{\Gamma}\frac{\rd{}^3\Gamma}{\rd{}\cos\theta_\PK\, \rd{}\cos\theta_l\, \rd{}q^2}  = \\
  \begin{aligned}
   \qquad&\frac{9}{16} \left\lbrace \left[ \frac{2}{3} F_S + \frac{4}{3} A_S \cos\theta_\PK \right] \left(1 - \cos^2\theta_l\right) \right. \\
    & \left. +\; \left(1 - F_S\right) \Bigl[2 F_L \cos^2\theta_\PK \left(1 - \cos^2\theta_l\right) \right. \\
    & \left. +\; \frac{1}{2} \left(1 - F_L\right) \left(1 - \cos^2\theta_\PK\right) \left(1 + \cos^2\theta_l\right) \right. \\
    & \left. +\; \frac{4}{3} A_\mathrm{FB} \left(1 - \cos^2\theta_\PK\right) \cos\theta_l\Bigr] \right\rbrace.
  \end{aligned}
\end{multline}
}{
\begin{equation}\label{eq:angALL}
  \begin{split}
     \frac{1}{\Gamma}\frac{\rd{}^3\Gamma}{\rd{}\!\cos\theta_\PK\, \rd{}\!\cos\theta_l\, \rd{}\!q^2}  &=
      \frac{9}{16} \left\lbrace \left[ \frac{2}{3} F_S + \frac{4}{3} A_S \cos\theta_\PK \right] \left(1 - \cos^2\theta_l\right) \right. \\
    & \left. + \left(1 - F_S\right) \Bigl[2 F_L \cos^2\theta_\PK \left(1 - \cos^2\theta_l\right) \right. \\
    & \left. + \frac{1}{2} \left(1 - F_L\right) \left(1 - \cos^2\theta_\PK\right) \left(1 + \cos^2\theta_l\right) \right. \\
    & \left. + \frac{4}{3} A_\mathrm{FB} \left(1 - \cos^2\theta_\PK\right) \cos\theta_l\Bigr] \right\rbrace.
   \end{split}
\end{equation}
}

The main results of the analysis are extracted from unbinned extended maximum-likelihood fits in
bins of $q^2$ to three variables: the $\PKp\Pgpm\Pgmp\Pgmm$ invariant mass and the two angular
variables ${\theta_\PK}$ and ${\theta_l}$.  For each $q^2$ bin, the probability density function (PDF)
has the following expression:
\ifthenelse{\boolean{cms@external}}{
\begin{multline} \label{eq:PDF}
\text{PDF}(m,\cos\theta_\PK,\cos\theta_l) = \\
\begin{aligned}
\qquad&Y_{S} \cdot S(m) \cdot S(\cos\theta_\PK,\cos\theta_l) \cdot \epsilon(\cos\theta_\PK,\cos\theta_l) \\
    & + Y_{Bc} \cdot B_{c}(m) \cdot B_{c}(\cos\theta_\PK) \cdot B_{c}(\cos\theta_l) \\
    & + Y_{Bp} \cdot B_{p}(m) \cdot B_{p}(\cos\theta_\PK) \cdot B_{p}(\cos\theta_l).
  \end{aligned}
\end{multline}
}{
\begin{equation} \label{eq:PDF}
\begin{split}
\text{PDF}(m,\cos\theta_\PK,\cos\theta_l) &=
     Y_{S} \cdot S(m) \cdot S(\cos\theta_\PK,\cos\theta_l) \cdot \epsilon(\cos\theta_\PK,\cos\theta_l) \\
    & + Y_{Bc} \cdot B_{c}(m) \cdot B_{c}(\cos\theta_\PK) \cdot B_{c}(\cos\theta_l) \\
    & + Y_{Bp} \cdot B_{p}(m) \cdot B_{p}(\cos\theta_\PK) \cdot B_{p}(\cos\theta_l).
  \end{split}
\end{equation}
}
The signal yield is given by the free parameter $Y_S$.  The signal shape is described by the product
of a function $S(m)$ of the invariant mass variable, the theoretical signal shape as a function of
two angular variables, $S(\cos\theta_\PK,\cos\theta_l)$, and the efficiency as a function of the same
two variables, $\epsilon(\cos\theta_\PK,\cos\theta_l)$.  The signal mass shape $S(m)$ is the sum of
two Gaussian functions with a common mean.  While the mean is free to float, the two resolution
parameters and the relative fraction are fixed to the result from a fit to the simulated events.
The signal angular function $S(\cos\theta_\PK,\cos\theta_l)$ is given by Eq.~(\ref{eq:angALL}). The
efficiency function $\epsilon(\cos\theta_\PK,\cos\theta_l)$, which also accounts for mistagging of a
$\PBz$ as a $\PaBz$ (and vice versa), is obtained by fitting the two-dimensional
efficiency histograms (6 $\cos\theta_\PK$ bins and 5 $\cos\theta_l$ bins) to polynomials in
$\cos\theta_\PK$ and $\cos\theta_l$.  The $\cos\theta_\PK$ polynomial is degree 3, while the
$\cos\theta_l$ polynomial is degree 6, with the 1st and 5th orders removed, as these were the simplest
polynomials that adequately described the efficiency in all bins.  For some $q^2$ bins, simpler
polynomials are used as they are sufficient to describe the data.  There are two contributions to
the background, with yields given by $Y_{Bp}$ for the ``peaking'' background and $Y_{Bc}$ for the
``combinatorial'' background.  The peaking background is due to the remaining \BtoKstJpsi and
\BtoKstpsip decays, not removed by the dimuon mass or $q^2$ requirements.  For these
events, the dimuon mass is reconstructed far from the true $\cPJgy$ or $\psi'$ mass, which results
in a reconstructed $\PBz$ mass similarly displaced from the true $\PBz$ mass.  The
shapes of this background in the mass, $B_{p}(m)$, and angular variables, $B_{p}(\cos\theta_\PK)$ and
$B_{p}(\cos\theta_l)$, are obtained from simulation of \BtoKstJpsi and \BtoKstpsip events, fit to
the sum of two Gaussian functions in mass and polynomials in $\cos{\theta_\PK}$ and $\cos{\theta_l}$.
The background yield is also obtained from simulation, properly normalized by comparing the
reconstructed \BtoKstJpsi and \BtoKstpsip yields in data and MC simulation.  The remaining
background, combinatorial in nature, is described by a single exponential in mass, $B_{c}(m)$, and a
polynomial in each angular variable, $B_{c}(\cos\theta_\PK)$ and $B_{c}(\cos\theta_l)$, varying
between degree 0 and 4, as needed to describe the data.

The results of the fit in each $q^2$ bin (including the $\cPJgy$ and $\psi'$ bins) are
$A_\mathrm{FB}$ and $F_L$.
In the fits to the data, the yield $Y_{Bp}$ and all but one of the parameters that define the shapes
of $S(m)$, $B_{p}(m)$, $B_{p}(\cos\theta_\PK)$, and $B_{p}(\cos\theta_l)$ are initially set to the values
obtained from simulation, with a Gaussian constraint defined by the uncertainty found in the fit to
the simulated events.  The $S(m)$ mass parameter is not constrained.
The first fit to
the data is to the control samples: \BtoKstJpsi and \BtoKstpsip.  The values for $F_S$ and
$A_S$ from the \BtoKstJpsi fit are used in the signal $q^2$ bins, with Gaussian constraints
defined by the uncertainties from the fit.  The longitudinal polarization fraction $F_L$ and the
scalar fraction $F_S$ are constrained to lie in the physical region of 0 to 1.  In addition, penalty
terms are added to ensure that $\left| A_\mathrm{FB} \right| < \frac{3}{4}\left(1-F_L\right)$ and $\left|
  A_S \right| < \frac{1}{2}\left[F_S + 3F_L\left(1-F_S\right)\right]$, which are necessary to avoid
a negative decay rate.

The differential branching fraction, $\rd{}\mathcal{B}/\rd{}q^2$, is measured relative to the
normalization channel \BtoKstJpsi using
\begin{equation} \label{eq:BF}
  \frac{\rd{}\mathcal{B}\left(\BtoKstmumu\right)}{\rd{}q^2} = \frac{Y_{S}}{Y_N} \frac{\epsilon_N}{\epsilon_{S}} \frac{\rd{}\mathcal{B}\left(\BtoKstJpsi\right)}{\rd{}q^2},
\end{equation}
where $Y_{S}$ and $Y_N$ are the yields of the signal and normalization channels, respectively,
$\epsilon_{S}$ and $\epsilon_N$ are the efficiencies of the signal and normalization channels,
respectively, and $\mathcal{B}\left(\BtoKstJpsi \right)$ is the world-average branching fraction for
the normalization channel~\cite{PDG}.  The yields are obtained with fits to the invariant-mass
distributions and the efficiencies are obtained by integrating over the angular variables using
the values obtained from the previously described fits.

Three methods are used to validate the fit formalism and results.  First, 1000 pseudo-experiment
samples are generated in each $q^2$ bin using the PDF in Eq.~(\ref{eq:PDF}).  The log-likelihood
values obtained from the fits to the data are consistent with the distributions from the
pseudo-experiments, indicating an acceptable goodness of fit.  The pull distributions obtained from
the pseudo-experiments indicate the uncertainties returned by the fit are generally overestimated by
0--10\%.  No attempt is made to correct the experimental uncertainties for this effect.  Second, a
fit is performed to a sample of MC simulation events that approximated the data sample in size and
composition.  The MC simulation sample contains a data-like mixture of four types of events.  Three
types of events are generated and simulated events from \BtoKstmumu, \BtoKstJpsi, and \BtoKstpsip
decays.  The last event type is the combinatorial background, which is generated based on the PDF in
Eq.~(\ref{eq:PDF}).  Third, the fit is performed on the normalization/control samples and the
results compared to the known values.  Biases observed from these three checks are treated as
systematic uncertainties, as described in Section~\ref{sec:Systematics}.

\section{Systematic uncertainties}
\label{sec:Systematics}

A variety of systematic effects are investigated and the impacts on the measurements of $F_L$,
$A_\mathrm{FB}$, and $\rd{}\mathcal{B}/\rd{}q^2$ are evaluated.

The finite sizes of the MC simulation samples used to measure the efficiency introduce a systematic
uncertainty of a statistical nature.  Alternative efficiency functions are created by randomly
varying the parameters of the efficiency polynomials within the fitted uncertainties for the MC
samples.  The alternative efficiency functions are applied to the data and the root-mean-squares of
the returned values taken as the systematic uncertainty.

The fit algorithm is validated by performing 1000 pseudo-experiments, generated and fit with the PDF
of Eq.~(\ref{eq:PDF}). The average deviation of the 1000 pseudo-experiments from the expected mean
is taken as the systematic uncertainty associated with possible bias from the fit algorithm.  This
bias is less than half of the statistical uncertainty for all measurements.  Discrepancies between
the functions used in the PDF and the true distribution can also give rise to biases.  To evaluate
this effect, a MC simulation sample similar in size and composition to the analyzed data set is fit
using the PDF of Eq.~(\ref{eq:PDF}).  The differences between the fitted values and the true values
are taken as the systematic uncertainties associated with the fit ingredients.

Mistagging a $\PBz$ as a $\PaBz$ (and vice versa) worsens the measured
$\PBz$ mass resolution.  A comparison of resolutions for data and MC simulations (varying the
mistag rates in the simulation) indicates the mistag rate may be as high as 12\%, compared to the
value of 8\% determined from simulation.  The systematic uncertainty in the mistag rate is
obtained from the difference in the final measurements when these two values are used.

The systematic uncertainty related to the contribution from the $\PK\Pgp$ S-wave (and interference
with the P-wave) is evaluated by taking the difference between the default results, obtained by
fitting with a function accounting for the S-wave (Eq.~(\ref{eq:angALL})), with the results from a
fit performed with no S-wave or interference terms ($F_S=A_S=0$ in Eq.~(\ref{eq:angALL})).

Variations of the background PDF shapes, versus mass and angles, are used to estimate the effect
from the choice of PDF shapes.  The mass-shape parameters of the peaking background, normally taken
from a fit to the simulation, are left free in the data fit and the difference adopted as a
systematic uncertainty.  The degree of the polynomials used to fit the angular shapes of the
combinatorial background are increased by one and the difference taken as a systematic uncertainty.
In addition, the difference in results obtained by fitting the mass-shape parameters using the data,
rather than using the result from simulations, is taken as the signal mass-shape systematic
uncertainty.

The effect of the experimental resolution of $\cos\theta_\PK$ and $\cos\theta_l$ is estimated as the
difference, when significant, of the returned values for $A_\mathrm{FB}$ and $F_L$ when the reconstructed
or generated values of $\cos\theta_\PK$ and $\cos\theta_l$ are used.  The effect of the dimuon mass
resolution is found to be negligible.

A possible difference between the efficiency computed with the simulation and the true efficiency
in data is tested by comparing the measurements of known observables between data and simulation
using the control channels.  The differences in the measurements of $F_L$ and $A_\mathrm{FB}$ are
computed using the \BtoKstJpsi decay.  For the differential branching fraction measurement, the
systematic uncertainty is estimated using the ratio of branching fractions
$\mathcal{B}(\BtoKstJpsimumu)/\mathcal{B}(\BtoKstpsipmumu)$, where our measured value of $15.5 \pm
0.4$ (statistical uncertainty only) is in agreement with the most-precise previously published value
of $16.2 \pm 0.5 \pm 0.3$~\cite{Aaij:2012dda}.  We use the difference of 4.3\% between these two
measurements as an estimate of the systematic uncertainty from possible $q^2$-dependent efficiency
mismodeling.

For the branching fraction measurement, a common normalization systematic uncertainty of 4.6\%
arises from the branching fractions of the normalization mode (\BtoKstJpsi and
$\cPJgy\to\Pgmp\Pgmm$)~\cite{PDG}.  Finally, variation of the number of pileup collisions is
found to have no effect on the results.

The systematic uncertainties are measured and applied in each $q^2$ bin, with the total systematic
uncertainty obtained by adding in quadrature the individual contributions.  A summary of the
systematic uncertainties is given in Table~\ref{tab:sys}; the ranges give the variation over the
$q^2$ bins.

\begin{table*}[bth]
  \centering
    \topcaption{Systematic uncertainty contributions for the measurements of $F_L$, $A_\mathrm{FB}$, and
      $\rd{}\mathcal{B}/\rd{}q^2$.  The $F_L$ and $A_\mathrm{FB}$ uncertainties are absolute values,
      while the $\rd{}\mathcal{B}/\rd{}q^2$ uncertainties are relative to the measured
      value.  The ranges given refer to the variations over the $q^2$ bins.\label{tab:sys}}
    \begin{tabular}{lD{,}{\text{--}}{2.3}D{,}{\text{--}}{2.3}D{,}{\text{--}}{2.3}}
      Systematic uncertainty & \multicolumn{1}{c}{$F_L \left(10^{-3}\right)$} & \multicolumn{1}{c}{$A_\mathrm{FB} \left(10^{-3}\right)$} & \multicolumn{1}{c}{$\rd{}\mathcal{B}/\rd{}q^2 (\%)$} \\
      \hline
      Efficiency statistical uncertainty          & 5 ,  7 &  3 , 5 & \multicolumn{1}{c}{1} \\
      Potential bias from fit algorithm           & 3 , 40 & 12 , 77 & 0 , 2.7 \\
      Potential bias from fit ingredients         & \multicolumn{1}{c}{0} & 0 , 17 & 0 , 7.1 \\
      Incorrect CP assignment of decay            & 2 , 6 & 2 , 6 & \multicolumn{1}{c}{0} \\
      Effect of $\PK\Pgp$ S-wave contribution     & 5 , 23 & 6 , 14 & \multicolumn{1}{c}{5} \\
      Peaking background mass shape               & 0 , 26 & 0 , 8 & 0 , 15 \\
      Background shapes vs.\ $\cos\theta_{L,K}$   & 3 , 180 & 4 , 160 & 0 , 3.3 \\
      Signal mass shape                           & \multicolumn{1}{c}{0} & \multicolumn{1}{c}{0} & \multicolumn{1}{c}{0.9} \\
      Angular resolution                          & 0 , 19 & \multicolumn{1}{c}{0} & \multicolumn{1}{c}{0} \\
      Efficiency shape                            & \multicolumn{1}{c}{16} & \multicolumn{1}{c}{4} & \multicolumn{1}{c}{4.3} \\
      Normalization to \BtoKstJpsi                & \multicolumn{1}{c}{---} & \multicolumn{1}{c}{---} & \multicolumn{1}{c}{4.6} \\
      \hline
      Total systematic uncertainty                & 31 , 190 & 18 , 180 & 8.6 , 17 \\
    \end{tabular}
\end{table*}

\section{Results}
\label{sec:Results}

\begin{figure}[hbtp]
  \begin{center}
    \includegraphics[width=0.48\textwidth]{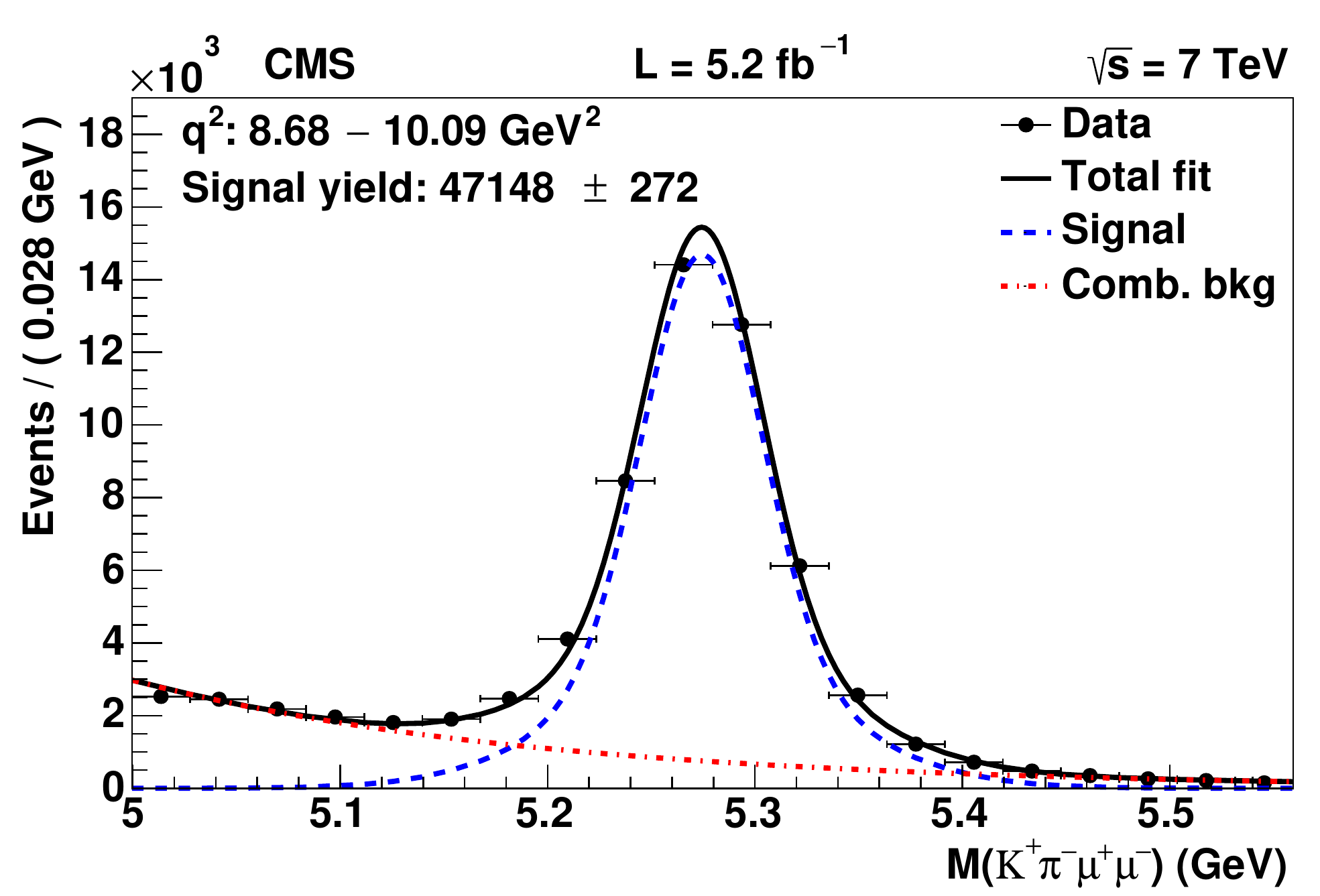}
    \includegraphics[width=0.48\textwidth]{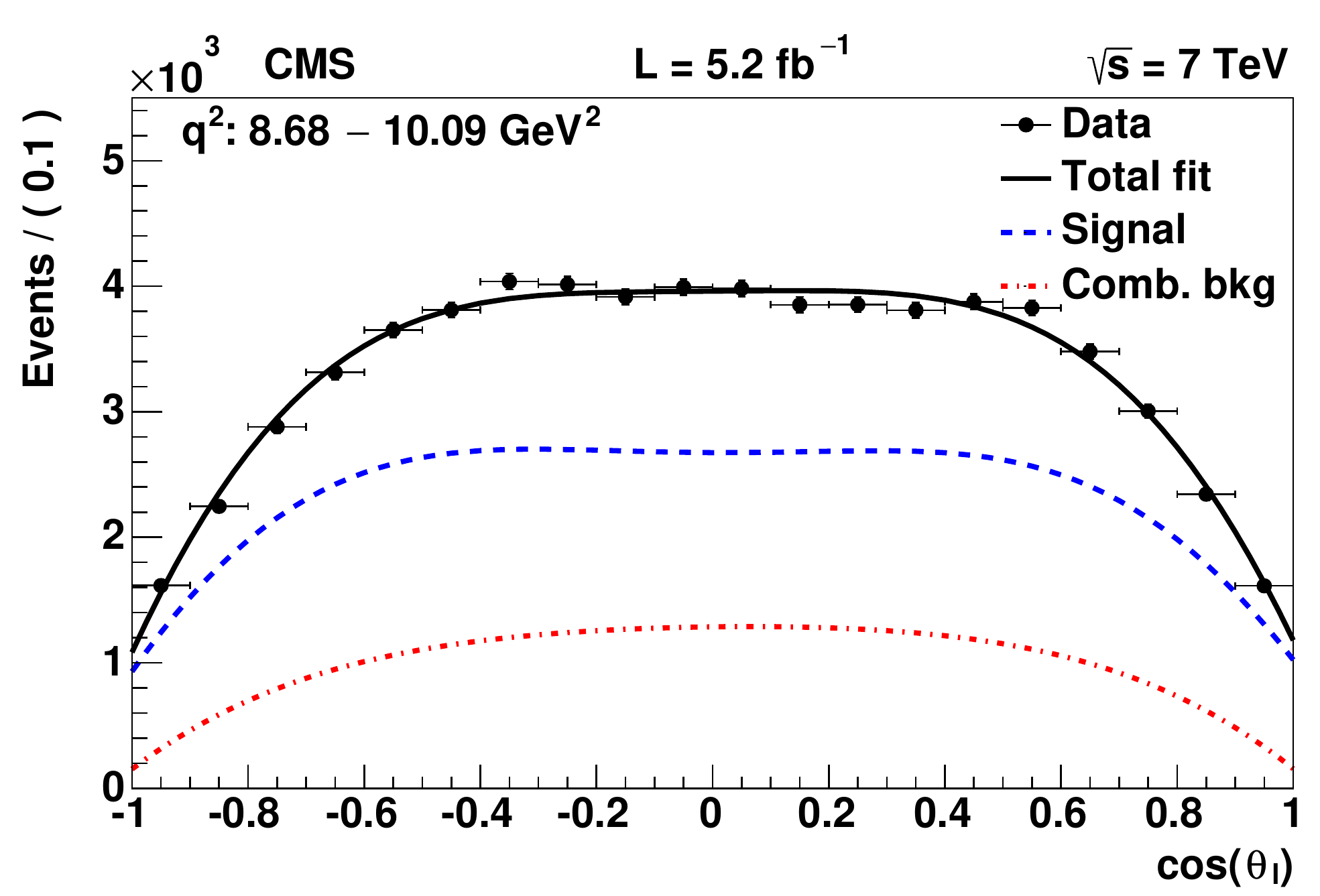}
    \includegraphics[width=0.48\textwidth]{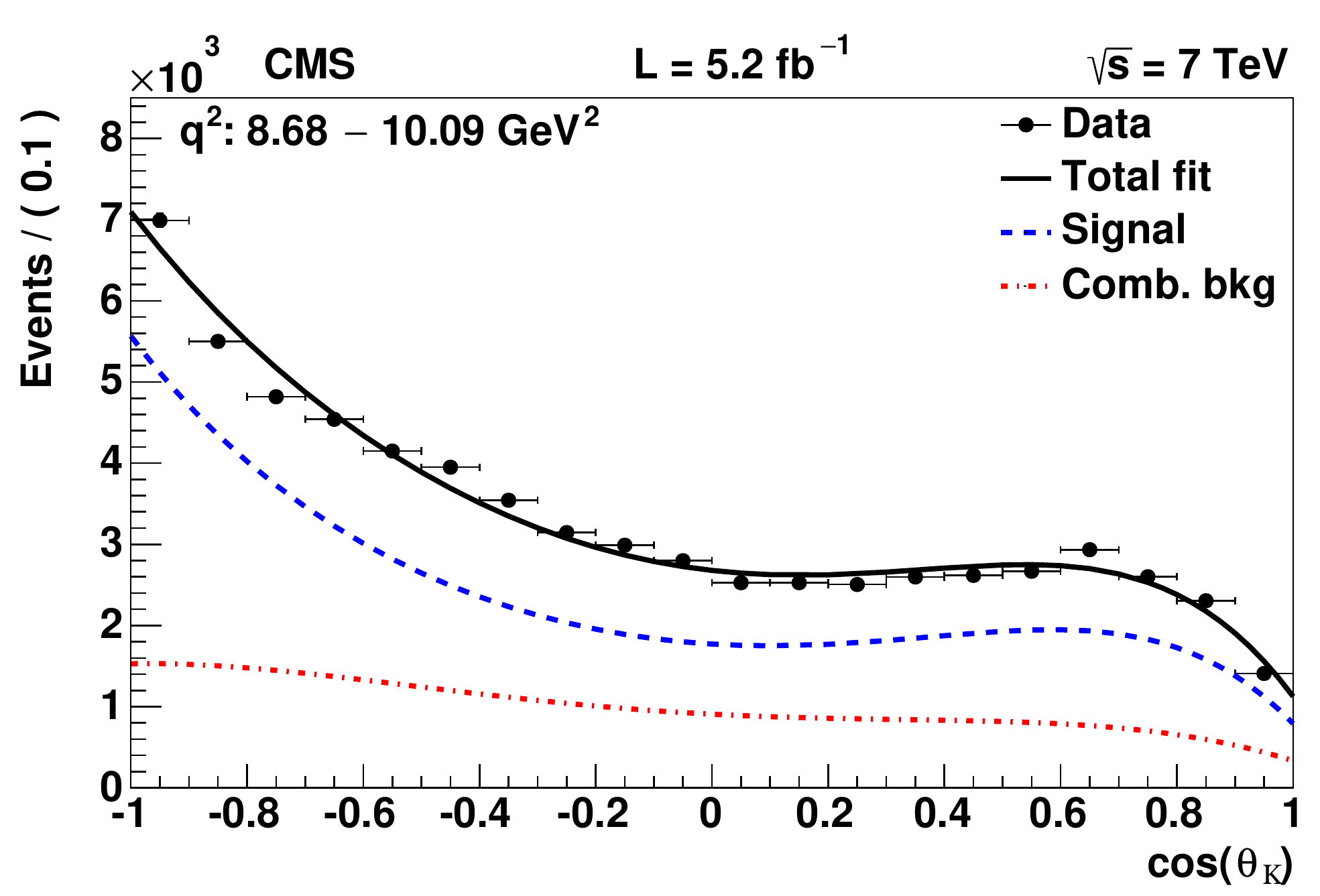}
    \caption{The $\PKp\Pgpm\Pgmp\Pgmm$ invariant-mass (\cmsTop), $\cos\theta_l$ (\cmsMiddle),
      and $\cos\theta_\PK$ (\cmsBottom) distributions for the $q^2$ bin associated with the \BtoKstJpsi
      decay, along with results from the projections of the overall unbinned maximum-likelihood fit
      (solid line), the signal contribution (dashed line), and the background contribution
      (dot-dashed line).}
  \label{fig:resPsiFlAfb}
  \end{center}
\end{figure}

The $\PKp\Pgpm\Pgmp\Pgmm$ invariant-mass, $\cos\theta_\PK$, and $\cos\theta_l$ distributions for
the $q^2$ bin corresponding to the \BtoKstJpsi decay are shown in Fig.~\ref{fig:resPsiFlAfb}, along
with the projection of the maximum-likelihood fit described in Section~\ref{sec:Analysis}.  The
results are used to validate the fitting procedure and obtain the values for $F_S$ and $A_S$ used in
the fits to the signal $q^2$ bins.  From 47\,000 signal events, the fitted values are $F_L =
0.554 \pm 0.004$, $A_\mathrm{FB} = -0.004 \pm 0.004$, $F_S = 0.01 \pm 0.01$, and $A_S = -0.10 \pm 0.01$,
where the uncertainties are statistical.  Considering also the typical systematic uncertainties
(Table~\ref{tab:sys}), the result for $F_L$ is compatible with the world-average value of $0.570 \pm
0.008$~\cite{PDG}, while the value for $A_\mathrm{FB}$ is consistent with the expected result of no
asymmetry.  The same fit is performed for the \BtoKstpsip $q^2$ bin, where 3200 signal events yield
results of $F_L = 0.509 \pm 0.016\stat$, which is consistent with the world-average value
of $0.46 \pm 0.04$~\cite{PDG}, and $A_\mathrm{FB} = 0.013 \pm 0.014\stat$, compatible with no
asymmetry, as expected in the SM\@.

\begin{figure*}[htbp]
  \begin{center}
    \includegraphics[width=0.48\textwidth]{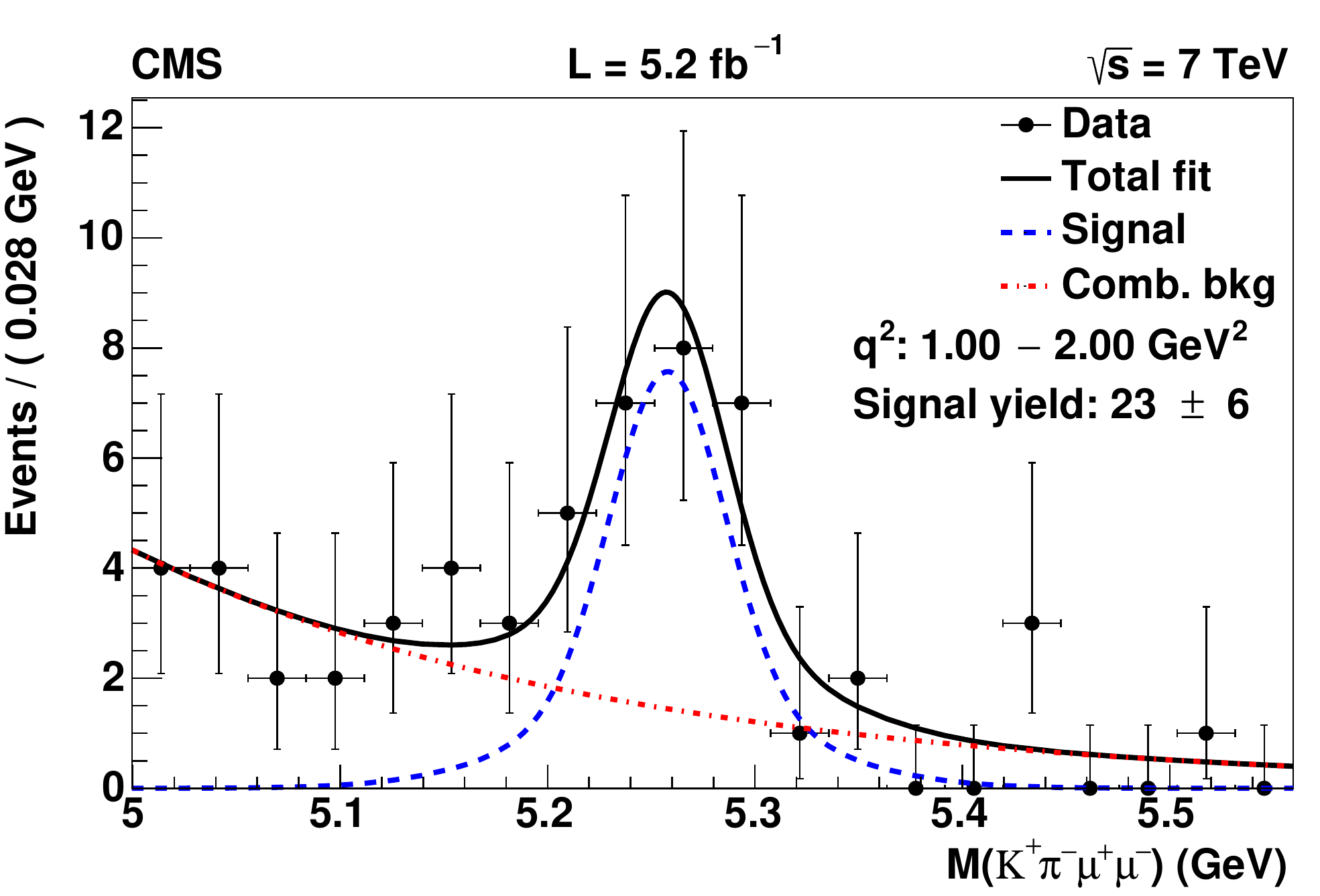}
    \includegraphics[width=0.48\textwidth]{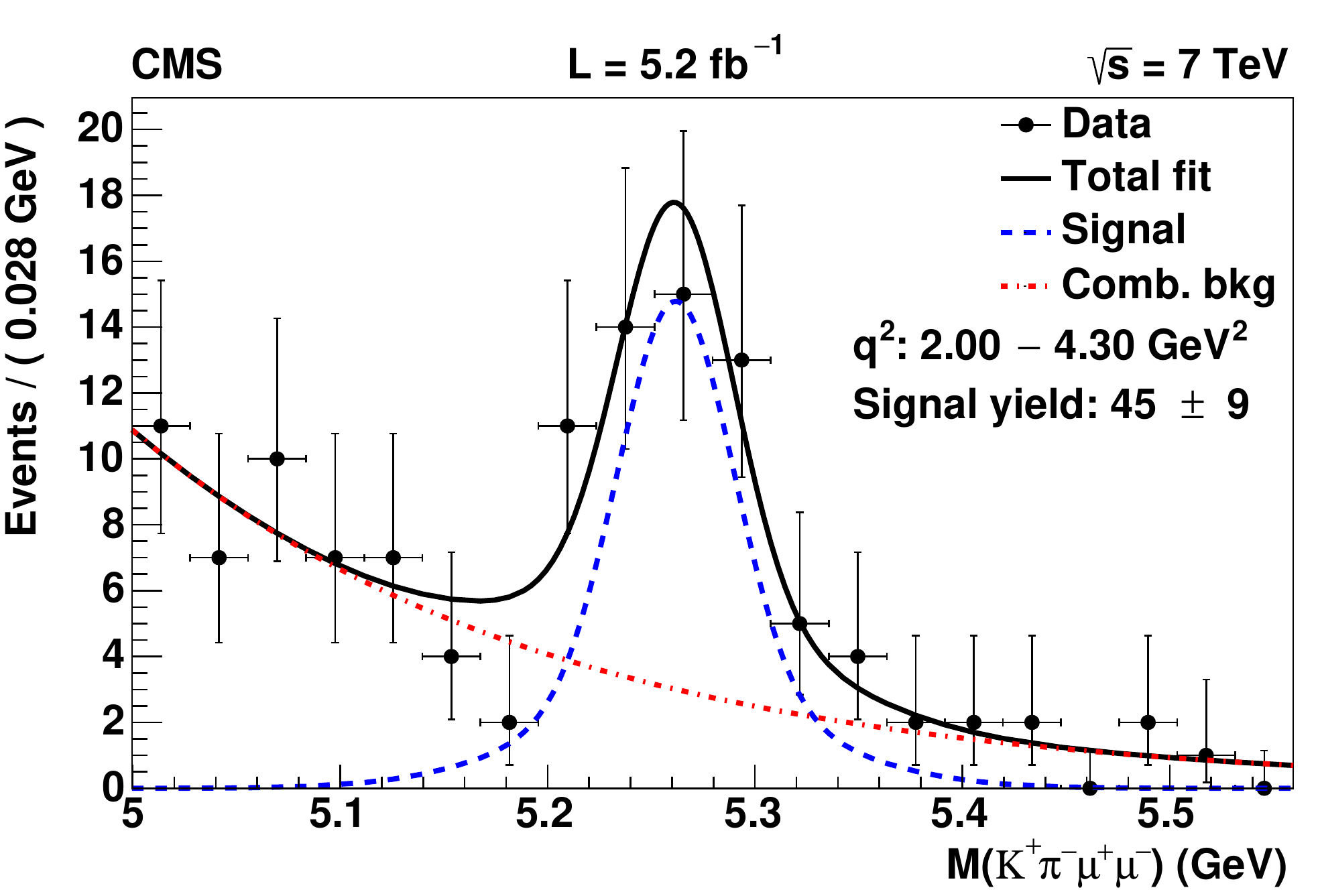}
    \includegraphics[width=0.48\textwidth]{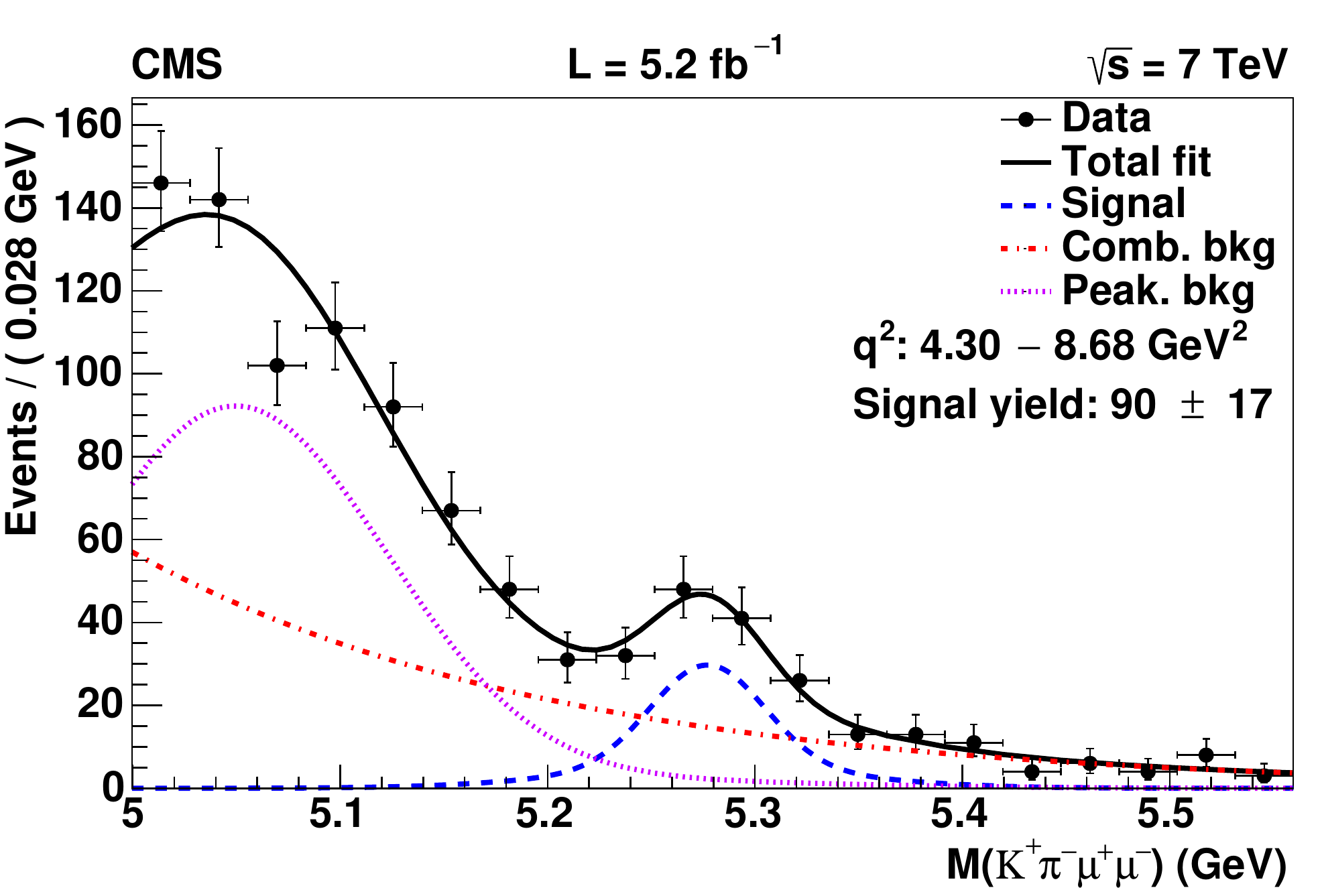}
    \includegraphics[width=0.48\textwidth]{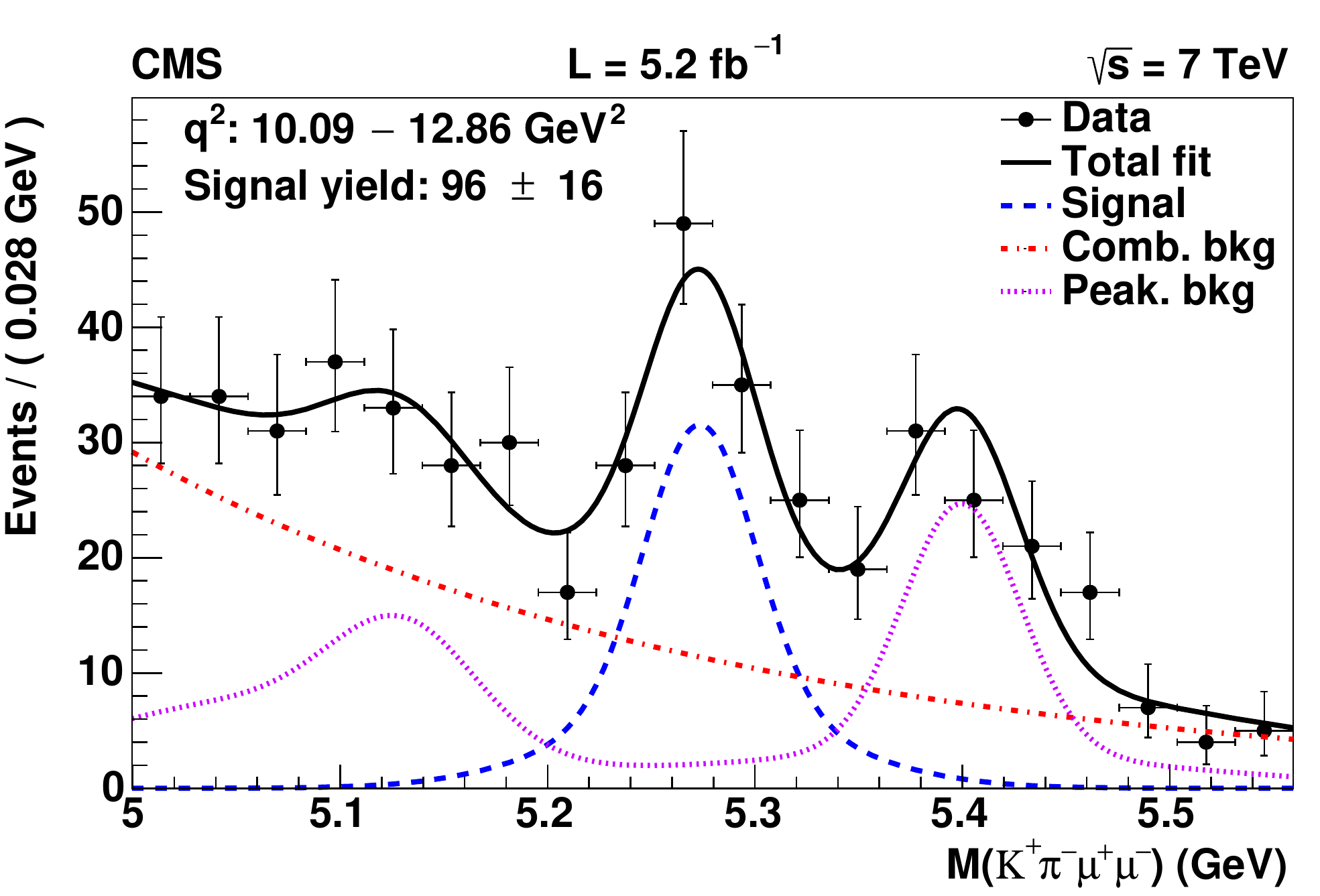}
    \includegraphics[width=0.48\textwidth]{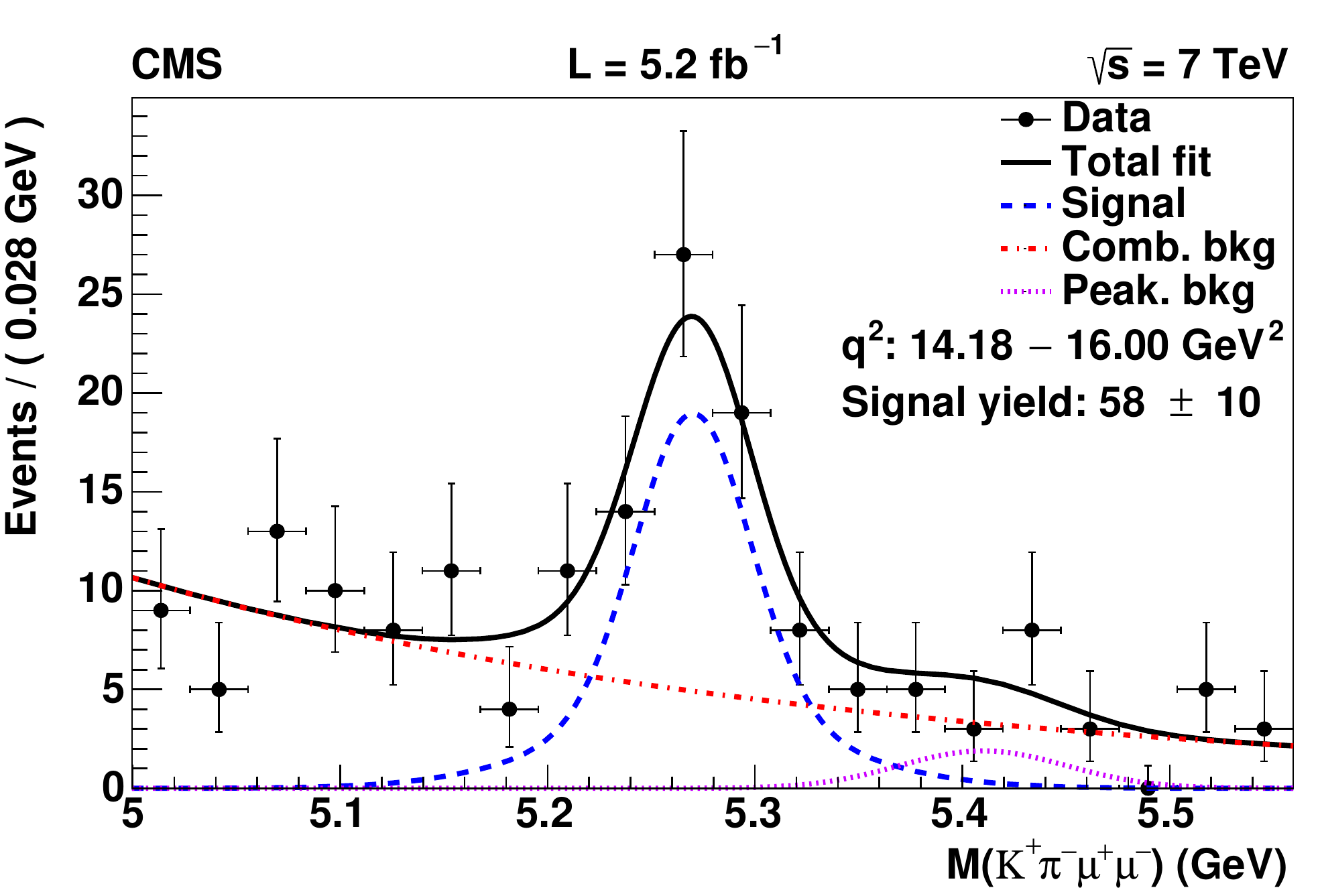}
    \includegraphics[width=0.48\textwidth]{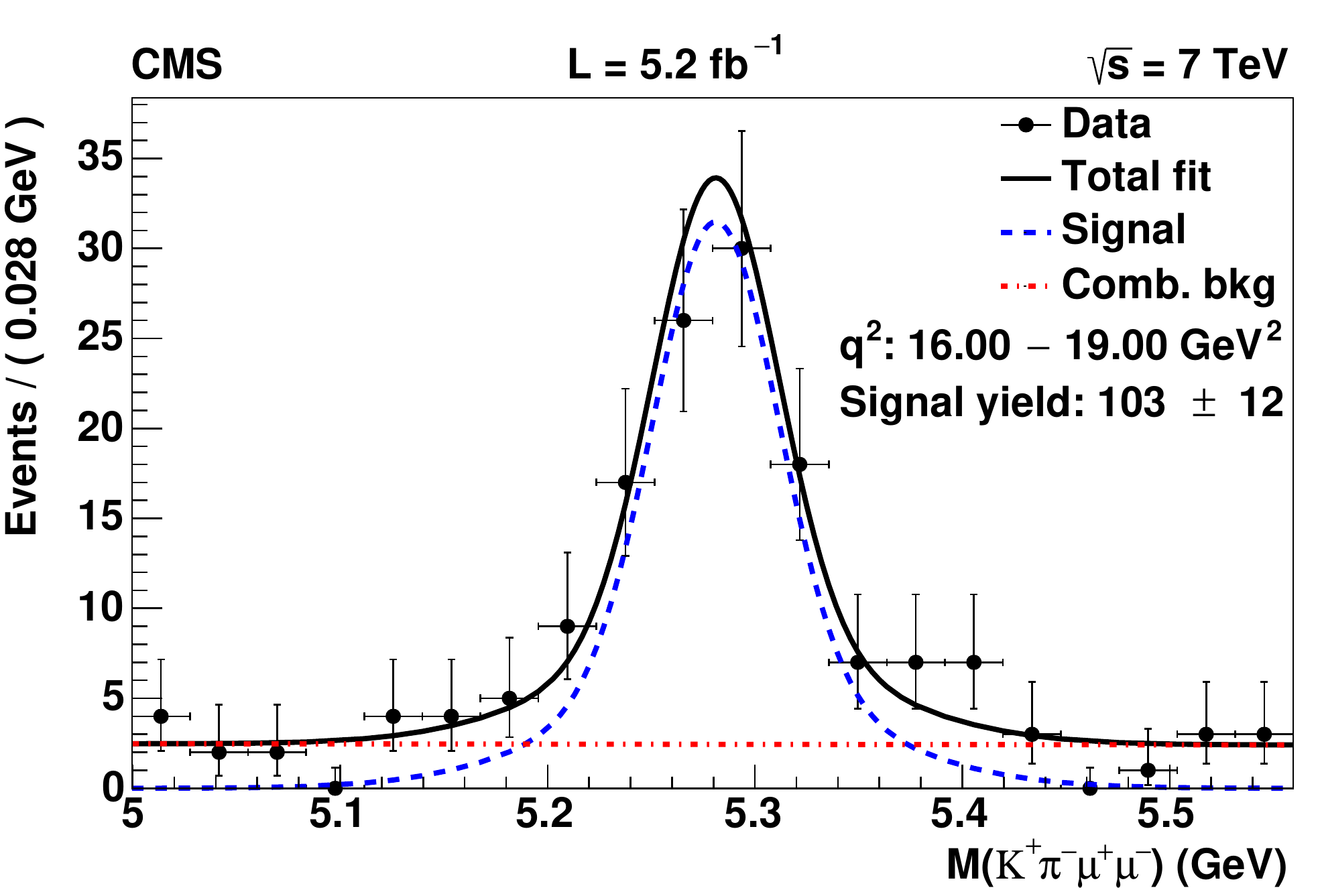}
    \caption{The $\PKp\Pgpm\Pgmp\Pgmm$ invariant-mass distributions for each of the signal
      $q^2$ bins.  Overlaid on each mass distribution is the projection of the unbinned
      maximum-likelihood fit results for the overall fit (solid line), the signal contribution
      (dashed line), the combinatorial background contribution (dot-dashed line), and the peaking
      background contribution (dotted line).}
    \label{fig:invMassq2}
  \end{center}
\end{figure*}

\begin{figure}[hbtp]
  \begin{center}
    \includegraphics[width=0.48\textwidth]{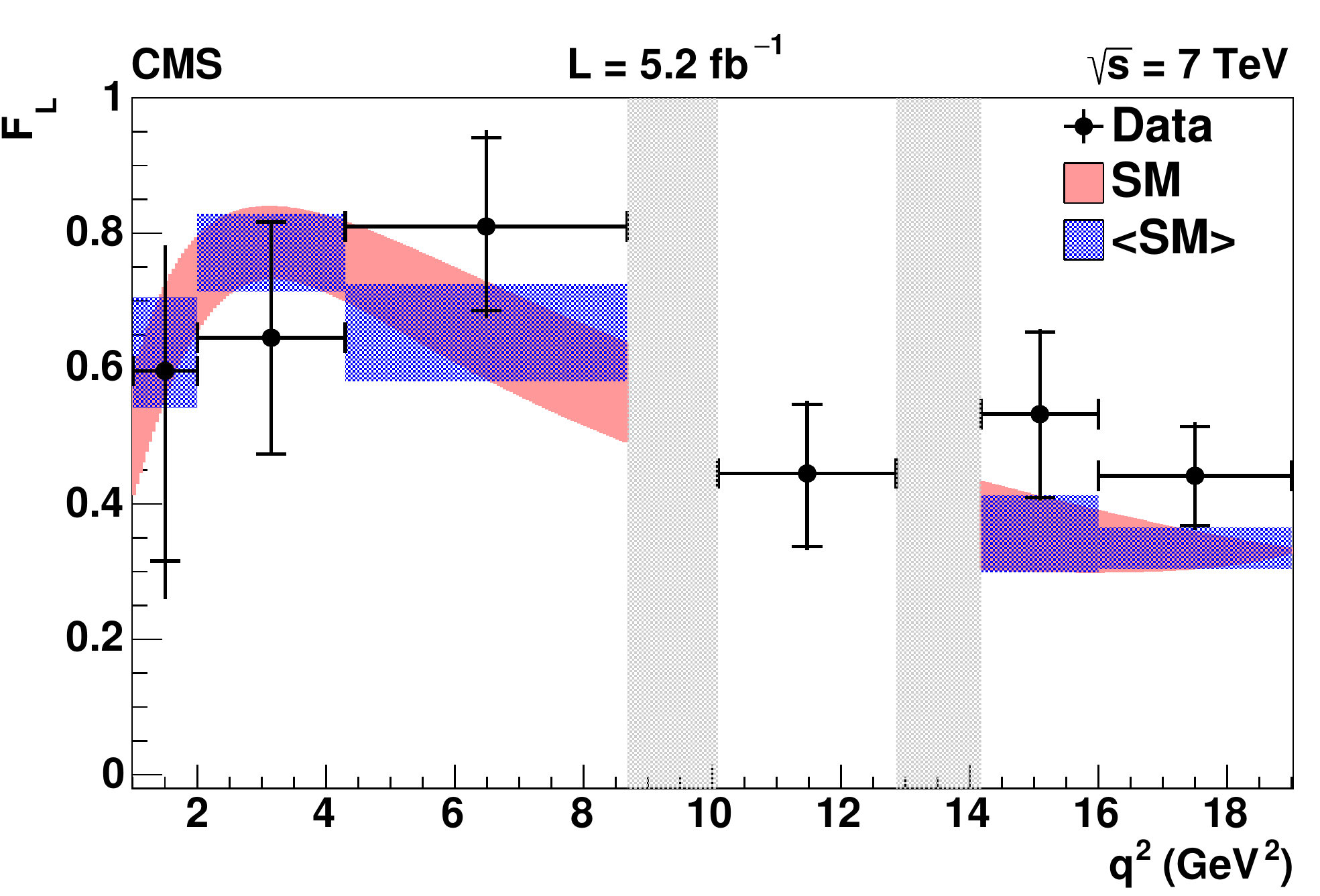}
    \includegraphics[width=0.48\textwidth]{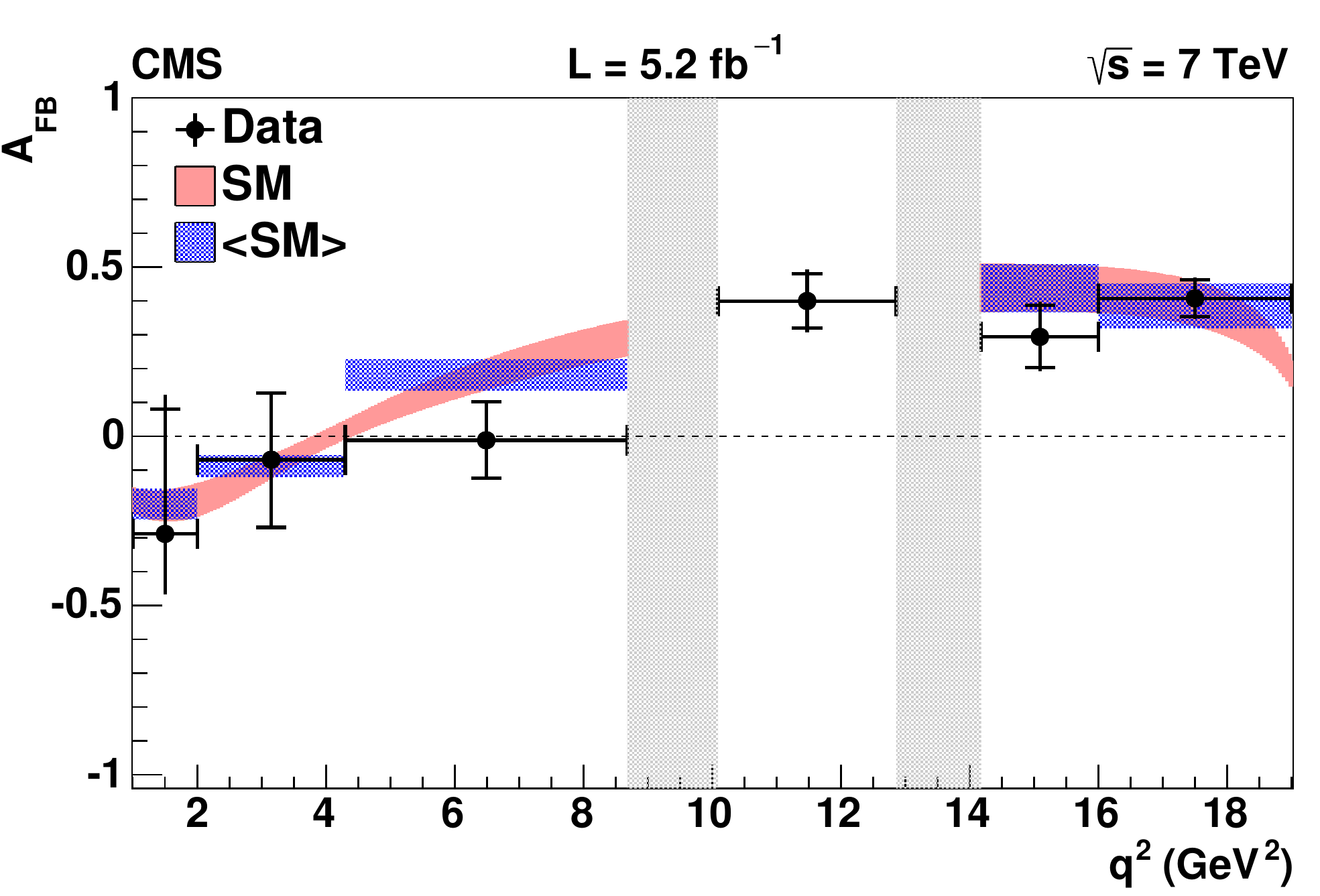}
    \caption{Results of the measurement of $F_L$ (\cmsLeft) and $A_\mathrm{FB}$ (\cmsRight) versus $q^2$.  The
      statistical uncertainty is shown by inner error bars, while the outer error bars give the
      total uncertainty.  The vertical shaded regions correspond to the $\cPJgy$ and $\psi'$
      resonances.  The other shaded regions show the SM prediction as a continuous distribution and
      after rate-averaging across the $q^2$ bins $(\langle \text{SM} \rangle)$ to allow direct
      comparison to the data points.  Reliable theoretical predictions between the $\cPJgy$ and
      $\psi'$ resonances $(10.09 < q^2 < 12.86 \GeV^2)$ are not available.}
    \label{fig:resultFLAFB}
  \end{center}
\end{figure}

The $\PKp\Pgpm\Pgmp\Pgmm$ invariant mass distributions for each $q^2$ bin of the signal sample
$\PBz\to\cPKstz$ $\Pgmp \Pgmm$ are shown in Fig.~\ref{fig:invMassq2},
along with the projection of the unbinned maximum-likelihood fit described in
Section~\ref{sec:Analysis}.  Clear signals are seen in each bin, with yields ranging from $23\pm 6$ to
$103\pm 12$ events.  The fitted results for $F_L$ and $A_\mathrm{FB}$ are
shown in Fig.~\ref{fig:resultFLAFB}, along with the SM predictions.  The values of $A_\mathrm{FB}$
and $F_L$ obtained for the first $q^2$ bin are at the physical boundary, which is enforced by a
penalty term.  This leads to statistical uncertainties, obtained from \textsc{minos}~\cite{Minuit},
of zero for the positive (negative) uncertainty for $F_L$ $(A_\mathrm{FB})$.

The SM predictions are taken from Ref.~\cite{Bobeth:2012vn} and combines two calculational
techniques. In the low-$q^2$ region, a QCD factorization approach~\cite{Beneke:2001at} is used,
which is applicable for $q^2<4m_c^4$, where $m_c$ is the charm quark mass.
In the high-$q^2$ region, an operator
product expansion in the inverse \cPqb-quark mass and $1/\!\sqrt{q^2}$~\cite{Grinstein:2004vb,Beylich:2011aq}
is combined with heavy quark form factor relations~\cite{Grinstein:2002cz}.  This is valid above the open-charm
threshold. In both regions, the
form factor calculations are taken from Ref.~\cite{Ball:2004rg}, and a dimensional estimate is made
of the uncertainty from the expansion corrections~\cite{Egede:2008uy}.  Other recent SM
calculations~\cite{Ali:2006ew,Altmannshofer:2011gn,Jager:2012uw,Descotes-Genon:2013vna} give similar
results, with the largest variations found in the uncertainty estimates and the differential branching fraction
value.  Between the $\cPJgy$ and $\psi'$ resonances, reliable theoretical predictions are not available.

\begin{figure}[btph]
  \begin{center}
    \includegraphics[width=0.48\textwidth]{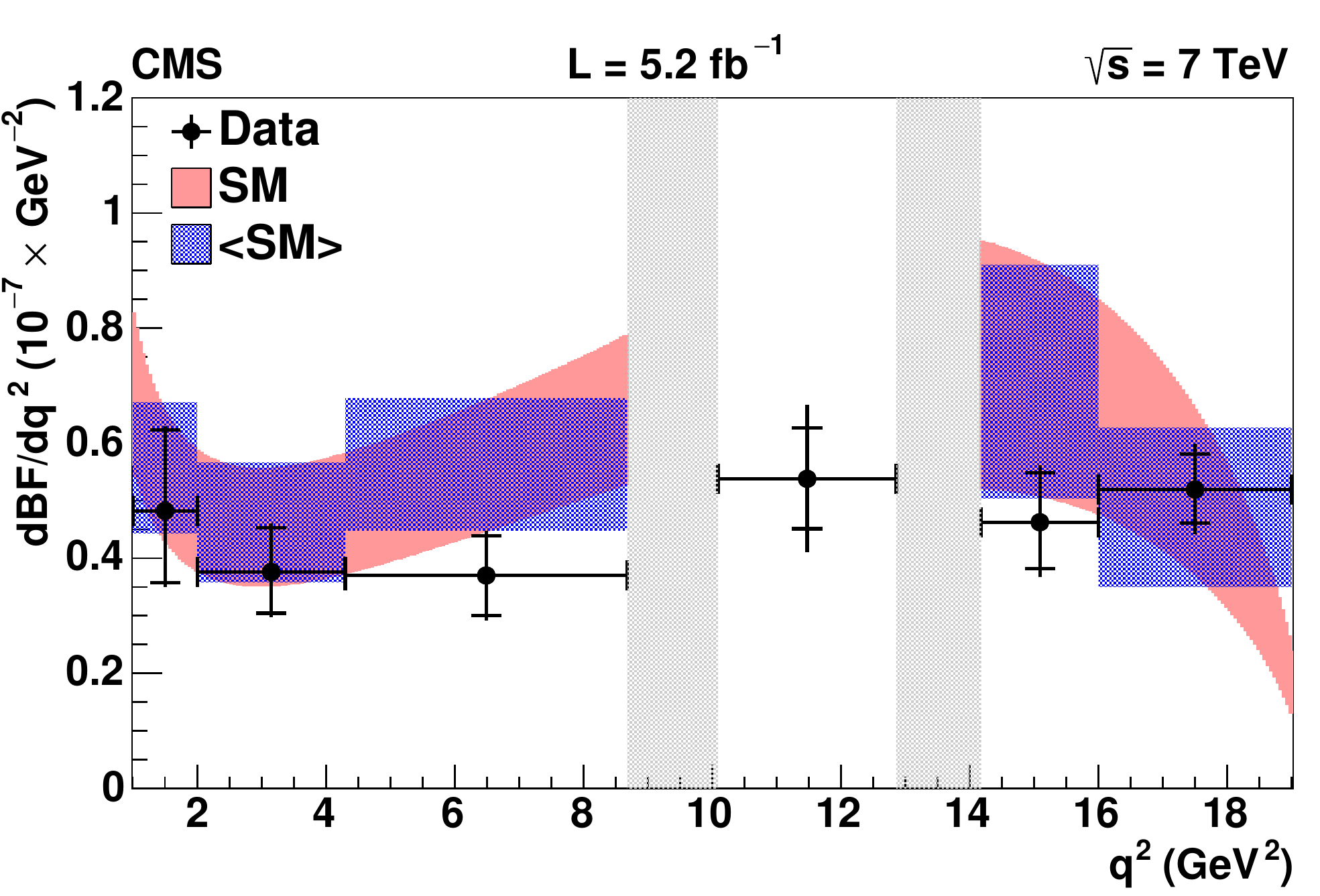}
    \caption{Results of the measurement of $\rd{}\mathcal{B}/\rd{}q^2$ versus $q^2$.  The
      statistical uncertainty is shown by inner error bars, while the outer error bars give the
      total uncertainty.  The vertical shaded regions correspond to the $\cPJgy$ and $\psi'$
      resonances.  The other shaded regions show the SM prediction as a continuous distribution and
      after rate-averaging across the $q^2$ bins $(\langle \text{SM} \rangle)$ to allow direct
      comparison to the data points.  Reliable theoretical predictions between the $\cPJgy$ and
      $\psi'$ resonances $(10.09 < q^2 < 12.86 \GeV^2)$ are not available.}
    \label{fig:resultBF}
  \end{center}
\end{figure}

Using the efficiency corrected yields for the signal and normalization modes (\BtoKstmumu and
\BtoKstJpsi) and the world-average branching fraction for the normalization mode~\cite{PDG}, the
branching fraction for \BtoKstmumu is obtained as a function of $q^2$, as shown in
Fig.~\ref{fig:resultBF}, together with the SM predictions.  The results for $A_\mathrm{FB}$, $F_L$, and
$\rd{}\mathcal{B}/\rd{}q^2$ are also reported in Table~\ref{tab:results}.

\begin{table*}[htbp]
\centering
  \topcaption{\label{tab:results}The yields and the measurements of $F_L$, $A_\mathrm{FB}$, and the branching
    fraction for the decay \BtoKstmumu in bins of $q^2$.  The first uncertainty is statistical and
    the second is systematic.}
\begin{tabular}{c|cccc}
$q^2$ & Yield & $F_L$  & $A_\mathrm{FB}$ & $\rd{}\mathcal{B}/\rd{}q^2$\\
$(\GeVns^2)$ & & &       & $(10^{-8}\GeV^{-2})$ \\[1pt]
\hline\\[-2ex]
1--2         & $23.0 \pm 6.3$ & $0.60^{\:+\:0.00}_{\:-\:0.28} \pm 0.19$ & $-0.29^{\:+\:0.37}_{\:-\:0.00} \pm 0.18$ & $4.8^{\:+\:1.4}_{\:-\:1.2} \pm 0.4$ \\[1pt]
2--4.3       & $45.0 \pm 8.8$ & $0.65 \pm 0.17 \pm 0.03$        & $-0.07\pm 0.20 \pm 0.02$         & $3.8\pm 0.7 \pm 0.3$ \\[1pt]
4.3--8.68    & $90 \pm 17$ & $0.81^{\:+\:0.13}_{\:-\:0.12} \pm 0.05$ & $-0.01\pm 0.11 \pm 0.03$         & $3.7\pm 0.7 \pm 0.4$ \\[1pt]
10.09--12.86 & $96 \pm 16$ & $0.45^{\:+\:0.10}_{\:-\:0.11} \pm 0.04$ & $0.40\pm 0.08 \pm 0.05$         & $5.4\pm 0.9 \pm 0.9$ \\[1pt]
14.18--16    & $58 \pm 10$ & $0.53 \pm 0.12 \pm 0.03$        & $0.29\pm 0.09 \pm 0.05$         & $4.6^{\:+\:0.9}_{\:-\:0.8} \pm 0.5$ \\[1pt]
16--19       & $103 \pm 12$ & $0.44 \pm 0.07 \pm 0.03$        & $0.41\pm 0.05 \pm 0.03$         & $5.2\pm 0.6 \pm 0.5$ \\[1pt]
\hline
1--6       & $107 \pm 14$ & $0.68 \pm 0.10 \pm 0.02$        & $-0.07\pm 0.12 \pm 0.01$         & $4.4\pm 0.6 \pm 0.4$ \\
\end{tabular}
\end{table*}

\begin{figure}[hbtp]
  \begin{center}
    \includegraphics[width=0.48\textwidth]{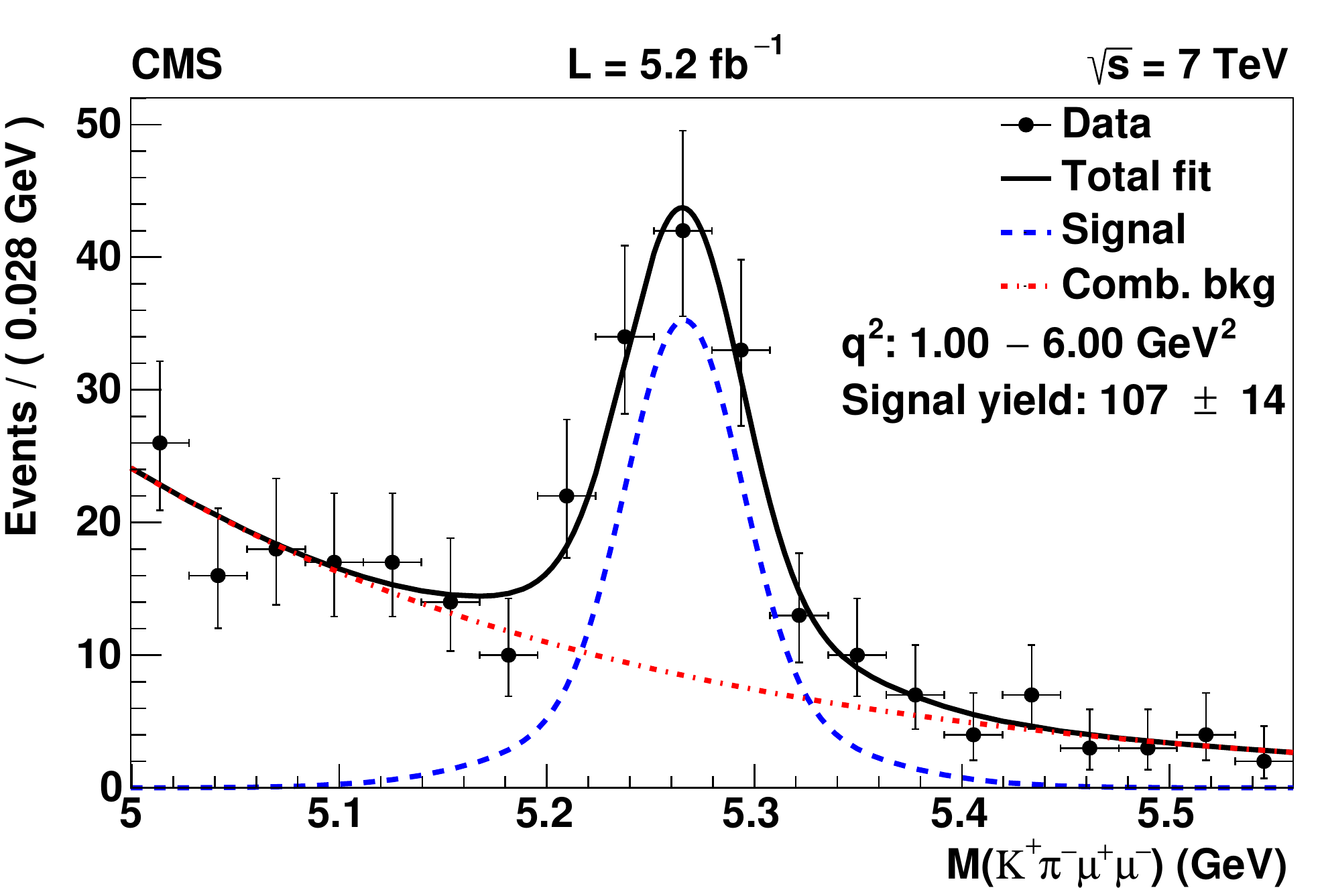}
    \includegraphics[width=0.48\textwidth]{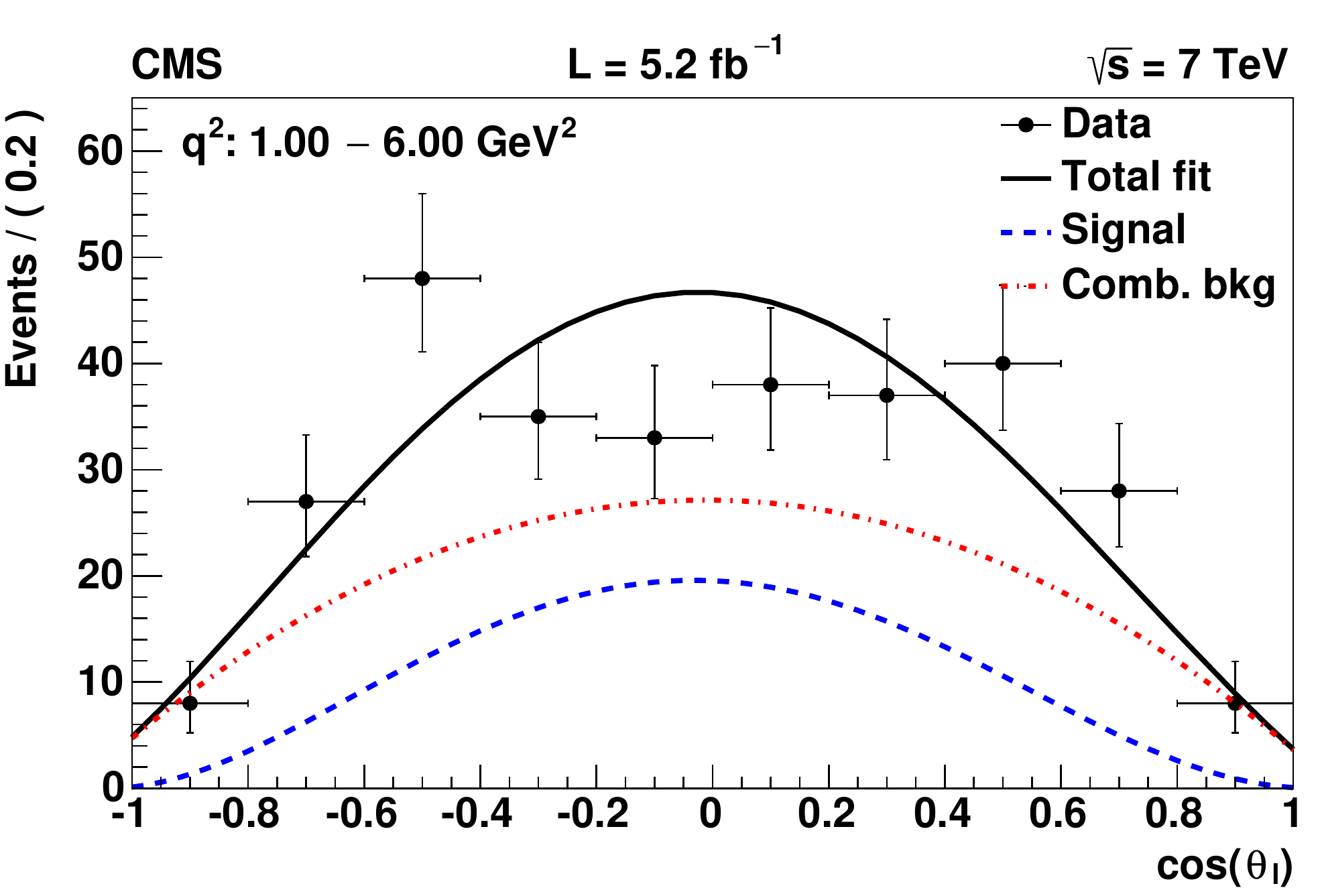}
    \includegraphics[width=0.48\textwidth]{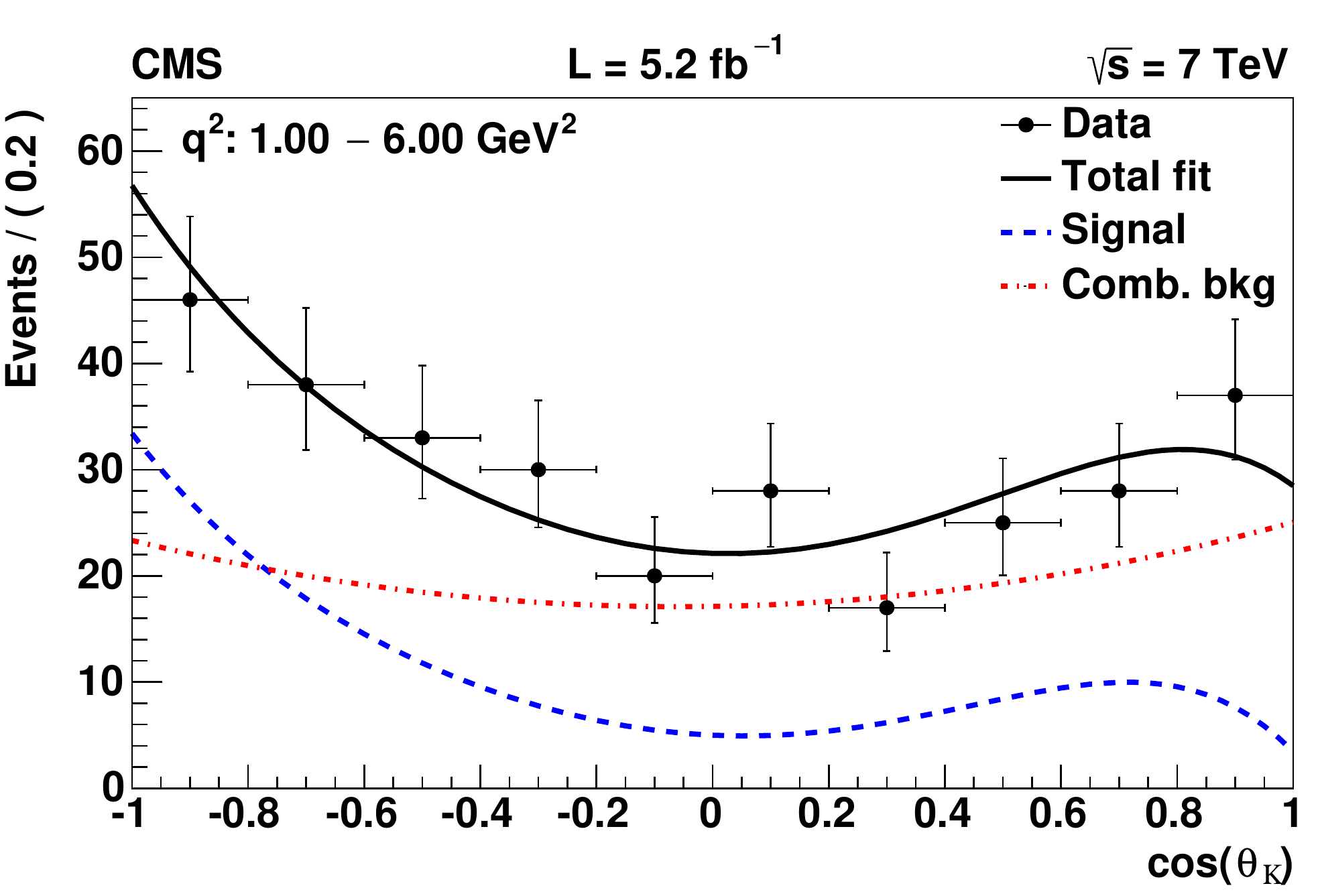}
    \caption{The $\PKp\Pgpm\Pgmp\Pgmm$ invariant-mass (\cmsTop), $\cos\theta_l$ (\cmsMiddle),
      and $\cos\theta_\PK$ (\cmsBottom) distributions for $1<q^2<6\GeV^2$, along with results from
      the projections of the overall unbinned maximum-likelihood fit (solid line), the signal
      contribution (dashed line), and the background contribution (dot-dashed line).}
  \label{fig:resSpec}
  \end{center}
\end{figure}

The angular observables can be theoretically predicted with good control of the relevant form-factor
uncertainties in the low dimuon invariant-mass region. It is therefore interesting to perform the
measurements of the relevant observables in the $1 < q^2 < 6\GeV^2$ region.  The experimental
results in this region, along with the fit projections, are shown in Fig.~\ref{fig:resSpec}.  The
values obtained from this fit for $F_L$, $A_\mathrm{FB}$, and  $\rd{}\mathcal{B}/\rd{}q^2$ are shown in
the bottom row of Table~\ref{tab:results}.
These results are consistent with the SM predictions of $F_L =
0.74^{\:+\:0.06}_{\:-\:0.07}$, $A_\mathrm{FB} = -0.05 \pm 0.03$, and $\rd{}\mathcal{B}/\rd{}q^2 =
(4.9^{\:+\:1.0}_{\:-\:1.1})\times 10^{-8}\GeV^{-2}$~\cite{Bobeth:2011gi}.

\begin{figure}[hbtp]
  \begin{center}
    \includegraphics[width=0.48\textwidth]{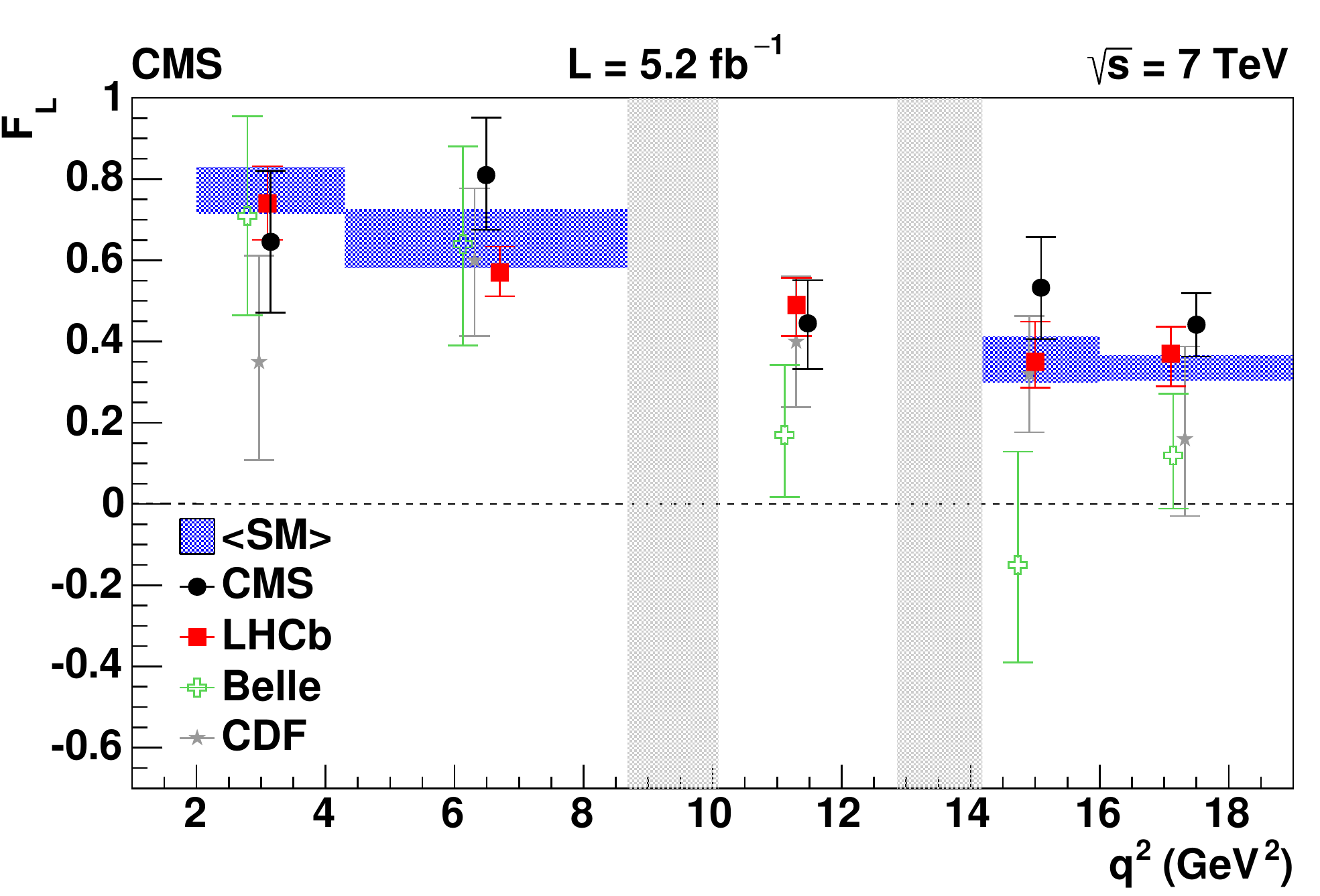}
    \includegraphics[width=0.48\textwidth]{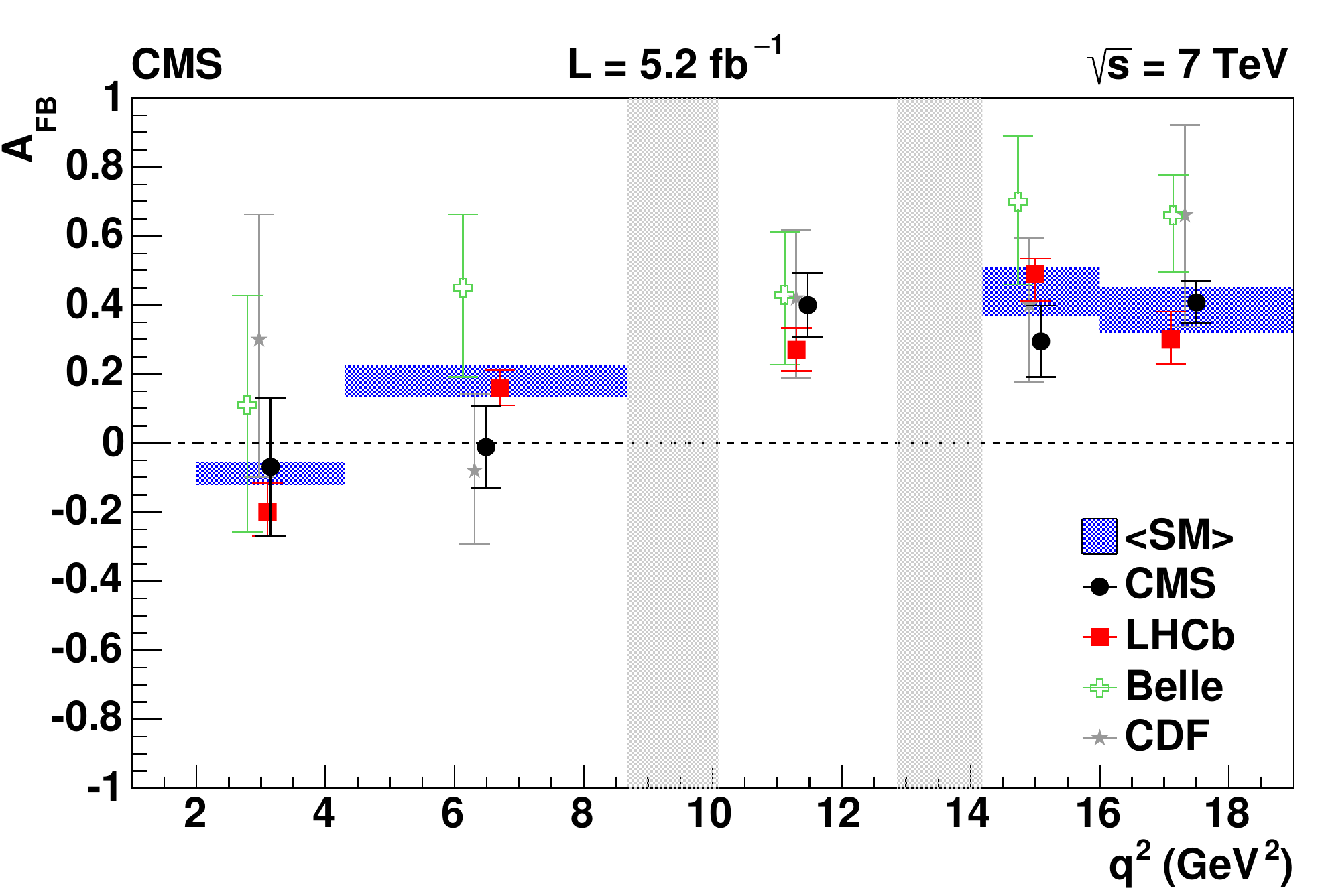}
    \includegraphics[width=0.48\textwidth]{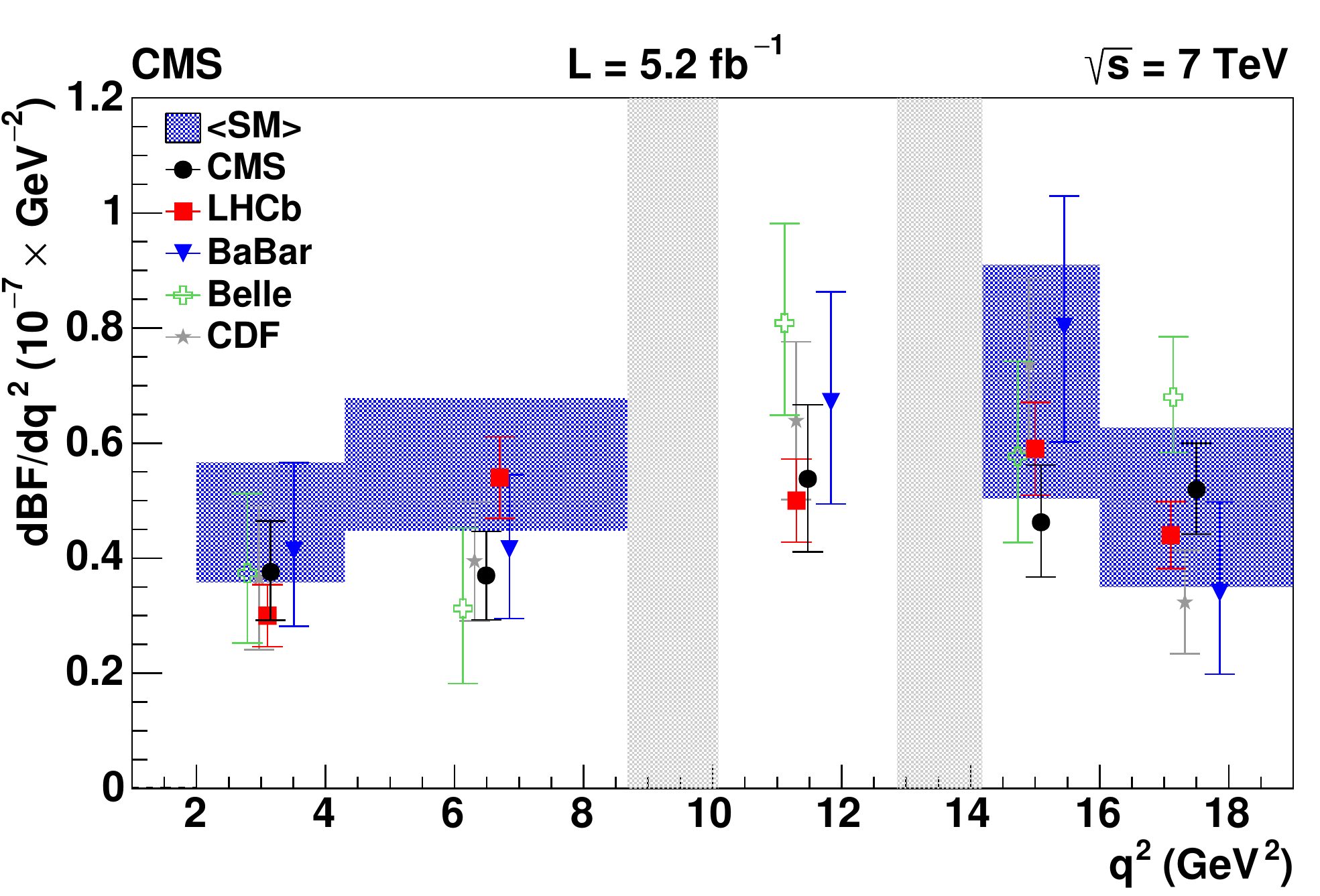}
    \caption{Measurements versus $q^2$ of $F_L$ (\cmsTop), $A_\mathrm{FB}$ (\cmsMiddle), and the branching
      fraction (\cmsBottom) for $\PB \to \PK^{\ast} \ell^+ \ell^-$ from CMS (this paper),
      Belle~\cite{Belle}, CDF~\cite{CDF,CDF_BR}, BaBar~\cite{BaBar_BR}, and LHCb~\cite{LHCb}.
      The error bars give the total uncertainty.  The vertical shaded regions correspond to the
      $\cPJgy$ and $\psi'$ resonances.  The other shaded regions are the result of rate-averaging
      the SM prediction across the $q^2$ bins to allow direct comparison to the data points.
      Reliable theoretical predictions between the $\cPJgy$ and $\psi'$ resonances $(10.09 < q^2 <
      12.86\GeV^2)$ are not available.}
    \label{fig:comp}
  \end{center}
\end{figure}

The results of $A_\mathrm{FB}$, $F_L$, and the branching fraction versus $q^2$ are compared to previous
measurements that use the same $q^2$ binning~\cite{Belle,CDF,CDF_BR,BaBar_BR,LHCb} in
Fig.~\ref{fig:comp}.  The CMS measurements are more precise than all but the LHCb values, and in the
highest-$q^2$ bin, the CMS measurements have the smallest uncertainty in $A_\mathrm{FB}$ and $F_L$.
Table~\ref{tab:comp} provides a comparison of the same quantities in the low dimuon invariant-mass
region: $1 < q^2 < 6\GeV^2$.

\begin{table*}[htbp]
\centering
  \topcaption{\label{tab:comp}Measurements from CMS (this paper), LHCb~\cite{LHCb},
    BaBar~\cite{BaBar_BR}, CDF~\cite{CDF,CDF_BR}, and Belle~\cite{Belle} of $F_L$, $A_\mathrm{FB}$, and
    $\rd{}\mathcal{B}/\rd{}q^2$ in the region $1 < q^2 < 6\GeV^2$ for the decay
    $\PB \to \cPKst \ell^+ \ell^-$.  The first uncertainty is statistical and the
    second is systematic. The SM predictions are also given~\cite{Bobeth:2012vn}.}
\begin{tabular}{c|cccc}
Experiment & $F_L$  & $A_\mathrm{FB}$ & $\rd{}\mathcal{B}/\rd{}q^2\;(10^{-8}\GeV^{-2})$ \\[1pt]
\hline
CMS      & $0.68 \pm 0.10 \pm 0.02$   & $-0.07\pm 0.12 \pm 0.01$         & $4.4\pm 0.6 \pm 0.4$ \\[1pt]
LHCb     & $0.65^{\:+\:0.08}_{\:-\:0.07}\pm 0.03$ & $-0.17 \pm 0.06 \pm 0.01$ & $3.4\pm 0.3^{\:+\:0.4}_{\:-\:0.5}$ \\[1pt]
BaBar    & --- & --- & $4.1^{\:+\:1.1}_{\:-\:1.0}\pm 0.1$ \\[1pt]
CDF      & $0.69^{\:+\:0.19}_{\:-\:0.21} \pm 0.08$ & $0.29^{\:+\:0.20}_{\:-\:0.23} \pm 0.07$         & $3.2\pm 1.1 \pm 0.3$ \\[1pt]
Belle    & $0.67 \pm 0.23 \pm 0.05$  & $0.26^{\:+\:0.27}_{\:-\:0.32} \pm 0.07$    & $3.0^{\:+\:0.9}_{\:-\:0.8} \pm 0.2$ \\[1pt]
\hline
SM       & $0.74^{\:+\:0.06}_{\:-\:0.07}$ & $-0.05 \pm 0.03$ & $4.9^{\:+\:1.0}_{\:-\:1.1}$ \\[1pt]
\end{tabular}
\end{table*}

\section{Summary}
\label{sec:End}

Using a data sample recorded with the CMS detector during 2011 and corresponding to an integrated
luminosity of 5.2\fbinv, an angular analysis of the decay \BtoKstmumu has been carried out. The
data used for this analysis include more than 400 signal decays and 50\,000 normalization/control
mode decays (\BtoKstJpsi and \BtoKstpsip).  Unbinned maximum-likelihood fits have been performed in
bins of the square of the dimuon invariant mass $(q^2)$ with three independent variables, the
$\PKp\Pgpm\Pgmp\Pgmm$ invariant mass and two decay angles, to obtain values of the
forward-backward asymmetry of the muons, $A_\mathrm{FB}$, and the fraction of longitudinal polarization of
the $\cPKstz$, $F_L$. Using these results, unbinned maximum-likelihood fits to the
$\PKp\Pgpm\Pgmp\Pgmm$ invariant mass in $q^2$ bins have been used to extract the differential
branching fraction $\rd{}\mathcal{B}/\rd{}q^2$. The results are consistent with the SM
predictions and previous measurements.  Combined with other measurements, these results can be used
to rule out or constrain new physics.

\section*{Acknowledgements}
We congratulate our colleagues in the CERN accelerator departments for the excellent performance of
the LHC and thank the technical and administrative staffs at CERN and at other CMS institutes for
their contributions to the success of the CMS effort. In addition, we gratefully acknowledge the
computing centres and personnel of the Worldwide LHC Computing Grid for delivering so effectively
the computing infrastructure essential to our analyses. Finally, we acknowledge the enduring support
for the construction and operation of the LHC and the CMS detector provided by the following funding
agencies: BMWF and FWF (Austria); FNRS and FWO (Belgium); CNPq, CAPES, FAPERJ, and FAPESP (Brazil);
MES (Bulgaria); CERN; CAS, MoST, and NSFC (China); COLCIENCIAS (Colombia); MSES (Croatia); RPF
(Cyprus); MoER, SF0690030s09 and ERDF (Estonia); Academy of Finland, MEC, and HIP (Finland); CEA and
CNRS/IN2P3 (France); BMBF, DFG, and HGF (Germany); GSRT (Greece); OTKA and NKTH (Hungary); DAE and
DST (India); IPM (Iran); SFI (Ireland); INFN (Italy); NRF and WCU (Republic of Korea); LAS
(Lithuania); CINVESTAV, CONACYT, SEP, and UASLP-FAI (Mexico); MBIE (New Zealand); PAEC (Pakistan);
MSHE and NSC (Poland); FCT (Portugal); JINR (Dubna); MON, RosAtom, RAS and RFBR (Russia); MESTD
(Serbia); SEIDI and CPAN (Spain); Swiss Funding Agencies (Switzerland); NSC (Taipei); ThEPCenter,
IPST, STAR and NSTDA (Thailand); TUBITAK and TAEK (Turkey); NASU (Ukraine); STFC (United Kingdom);
DOE and NSF (USA).

Individuals have received support from the Marie-Curie programme and the European Research Council
and EPLANET (European Union); the Leventis Foundation; the A. P. Sloan Foundation; the Alexander von
Humboldt Foundation; the Belgian Federal Science Policy Office; the Fonds pour la Formation \`a la
Recherche dans l'Industrie et dans l'Agriculture (FRIA-Belgium); the Agentschap voor Innovatie door
Wetenschap en Technologie (IWT-Belgium); the Ministry of Education, Youth and Sports (MEYS) of Czech
Republic; the Council of Science and Industrial Research, India; the Compagnia di San Paolo
(Torino); the HOMING PLUS programme of Foundation for Polish Science, cofinanced by EU, Regional
Development Fund; and the Thalis and Aristeia programmes cofinanced by EU-ESF and the Greek NSRF.

\bibliography{auto_generated}
\cleardoublepage \appendix\section{The CMS Collaboration \label{app:collab}}\begin{sloppypar}\hyphenpenalty=5000\widowpenalty=500\clubpenalty=5000\textbf{Yerevan Physics Institute,  Yerevan,  Armenia}\\*[0pt]
S.~Chatrchyan, V.~Khachatryan, A.M.~Sirunyan, A.~Tumasyan
\vskip\cmsinstskip
\textbf{Institut f\"{u}r Hochenergiephysik der OeAW,  Wien,  Austria}\\*[0pt]
W.~Adam, T.~Bergauer, M.~Dragicevic, J.~Er\"{o}, C.~Fabjan\cmsAuthorMark{1}, M.~Friedl, R.~Fr\"{u}hwirth\cmsAuthorMark{1}, V.M.~Ghete, N.~H\"{o}rmann, J.~Hrubec, M.~Jeitler\cmsAuthorMark{1}, W.~Kiesenhofer, V.~Kn\"{u}nz, M.~Krammer\cmsAuthorMark{1}, I.~Kr\"{a}tschmer, D.~Liko, I.~Mikulec, D.~Rabady\cmsAuthorMark{2}, B.~Rahbaran, C.~Rohringer, H.~Rohringer, R.~Sch\"{o}fbeck, J.~Strauss, A.~Taurok, W.~Treberer-Treberspurg, W.~Waltenberger, C.-E.~Wulz\cmsAuthorMark{1}
% ...
\vskip\cmsinstskip
% ...
1:~~Also at Vienna University of Technology, Vienna, Austria\\
2:~~Also at CERN, European Organization for Nuclear Research, Geneva, Switzerland\\
% ...

\end{sloppypar}
\end{document}